\documentclass[manuscript,screen]{acmart}

\usepackage{soul}
\usepackage{url}
\usepackage[utf8]{inputenc}
\usepackage{graphicx}
\usepackage{algorithm}
\usepackage{algpseudocode}
\usepackage{adjustbox}
\usepackage{subfigure}

\AtBeginDocument{%
  \providecommand\BibTeX{{%
    \normalfont B\kern-0.5em{\scshape i\kern-0.25em b}\kern-0.8em\TeX}}}

\setcopyright{acmcopyright}
\copyrightyear{2018}
\acmYear{2018}
\acmDOI{10.1145/1122445.1122456}

\acmConference[Woodstock '18]{Woodstock '18: ACM Symposium on Neural
  Gaze Detection}{June 03--05, 2018}{Woodstock, NY}
\acmBooktitle{Woodstock '18: ACM Symposium on Neural Gaze Detection,
  June 03--05, 2018, Woodstock, NY}
\acmPrice{15.00}
\acmISBN{978-1-4503-XXXX-X/18/06}

\begin{document}

\title{ABEM: An Adaptive Agent-based Evolutionary Approach for Mining Influencers in Online Social Networks}


\author{Weihua Li}
\email{weihua.li@aut.ac.nz}
\affiliation{
  \institution{Auckland University of Technology}
  \city{Auckland}
  \country{New Zealand}
}

\author{Yuxuan Hu}
\email{yuxuan.hu@utas.edu.au}
\affiliation{
  \institution{University of Tasmania}
  \city{Hobart}
  \country{Australia}
}

\author{Shiqing Wu}
\email{shiqing.wu@utas.edu.au}
\affiliation{
  \institution{University of Tasmania}
  \city{Hobart}
  \country{Australia}
}

\author{Quan Bai}
\email{quan.bai@utas.edu.au}
\affiliation{
  \institution{University of Tasmania}
  \city{Hobart}
  \country{Australia}
}

\author{Edmund Lai}
\email{edmund.lai@aut.ac.nz}
\affiliation{
  \institution{Auckland University of Technology}
  \city{Auckland}
  \country{New Zealand}
}

\renewcommand{\shortauthors}{Li, et al.}

\begin{abstract}
A key step in influence maximization in online social networks is the identification of a small number of users, known as influencers, who are able to spread influence quickly and widely to other users.
The evolving nature of the topological structure of these networks makes it difficult to locate and identify these influencers. 
In this paper, we propose an adaptive agent-based evolutionary approach to address this problem in the context of both static and dynamic networks.
This approach is shown to be able to adapt the solution as the network evolves. It is also applicable to large-scale networks due to its distributed framework. 
Evaluation of our approach is performed by using both synthetic networks and real-world datasets. 
Experimental results demonstrate that the proposed approach outperforms state-of-the-art seeding algorithms in terms of maximizing influence.
\end{abstract}




\maketitle

\section{Introduction} 
In recent years, influence diffusion modelling and maximization in complex networks such as online social networks have attracted great deal of attention from both researchers and practitioners. 
It has applications to many areas, including decision making, marketing, and social influence analysis~\cite{kempe2003maximizing,zhou2014preference,sumith2018influence,zhang19understanding,Ankita2019location}. 
Influence maximization is achieved through the identification of a small number of users who are capable of spreading their influence quickly and widely, thus maximizing the impact across entire the social network \cite{kempe2003maximizing,sumith2018influence}.  
The process of influencer mining, known as seed selection, to obtain a set of influencers, known as a seed set, is a very challenging problem for a large network.
This is because users join and quit, the relationships form and vanish, and the strength of these relationships varies over time.
Thus the topology of the network evolves continuously \cite{ding2020influence}. 

Traditional diffusion models, such as the Independent Cascade (IC) model 
and the Linear Threshold (LT) model \cite{kempe2003maximizing} 
, and more recent seeding algorithms \cite{ju2020new,wang2018maximizing} 
are not applicable as they assume static network topologies. 
Specifically, most classic diffusion models require a global view of the entire social network to initiate influence propagation. The seeding algorithms require re-calculation if any network changes occur. 
A work-around is to utilize a large number of static snapshots.
However, this approach inevitably creates another set of big data \cite{li2018siminer}.  
Furthermore, such an approach turns out to be inefficient since the heuristics identified from the previous time steps are not utilized, but explores the solutions from the scratch. 
An alternative approach is to view influence maximization 
as an optimization problem within a dynamic, large-scale and distributed space. 
Evolutionary computation has been widely adopted to solve a wide range of optimization problems, and Genetic Algorithm (GA) is acknowledged as one of the typical evolutionary computation approaches \cite{grefenstette1985genetic,panizo2020multi}. GA facilitates an efficient searching process in a well-defined problem space, which incorporates a massive number of encoded candidate solutions, referring to as seed sets in the influence maximization problem. 
On the other hand, Agent-based Modelling (ABM) demonstrates a great many advantages in modelling complex systems, which has been widely adopted to explore the macro world through defining the micro-level of the system~\cite{bonabeau2002agent,macal2009agent,chen2008large}. 
With the traditional approaches, such as IC model and LT model, influence propagation demonstrates a hopping and infecting process. 
In other words, the influence is initiated from the active nodes, hopping further away via the network topologies. By contrast, using a Multi-Agent System (MAS) to model the spread of influences can present an evolutionary pattern of a network driven by individuals' actions, which enables the dynamic features to be captured over time  \cite{li2019amultiagent}.

In this paper, we propose an Adaptive Agent-Based Evolutionary Model (ABEM) by amalgamating both GA and ABM to tackle the influence maximization problem. The advantages of our proposed model are reflected as threefold. \textbf{First}, in the proposed model, ABM produces a reasonable search scope and optimizes the problem space through defining the individual agent's behaviours. As GA cannot guarantee the most optimal solution and the quality also deteriorates with the increase of problem space, ABM mitigates the limitations of GA by narrowing down the searching space. Specifically, before joining the problem space, agents (users) interact with neighbours and conduct initial evaluations locally, which significantly reduces the number of candidate solutions. \textbf{Second}, ABM enables GA to construct, evaluate and select solutions within a dynamic problem space. In the current setting, the network topological structure evolves over time. We utilize an influencer pool, formed through the agents' nomination behaviours, i.e., proposing them as the influencers and joining the influencer pool. With the evolution of the network, the elements of influencer pool also varies accordingly, providing an up-to-date problem space for influence maximization. \textbf{Third}, the proposed ABEM is also applicable in a large-scale environment, since the computation cost of the influence evaluation distributes to the individual agent. On top of that, we conduct four experiments to evaluate the performance of ABEM in both static and dynamic environments by using real-world datasets. We also further explore the adaptation behaviours of ABEM and conduct parameter analysis. The experimental results demonstrate the advantages of ABEM in mining influencers in static and dynamic social networks. To summarize, our main contributions of this research work can be reflected in three-fold: 

\begin{itemize}
	\item We first utilized a distributed approach with evolutionary algorithms to address the influence maximization problem. 
	\item We proposed a novel approach, which can handle large-scale social networks by distributing the major computational cost to the individual but retaining the optimization process in a central component. 
	\item We developed a novel mechanism for mining influencers with adaptation capabilities, which can handle the fast-changing environment of online social networks. 
\end{itemize}


The rest of this paper is organized as follows. In Section \ref{sec:relatedWorks}, we review the related work. The preliminaries and formal definitions are given in Section \ref{sec:preliminaries}. In Section \ref{sec:ABEM}, we demonstrate the overall process and provide a detailed description of the proposed ABEM approach. Experiments and the analysis of experimental results are presented in Section \ref{sec:experiments}. The paper is concluded in Section \ref{sec:conclusion}.

\section{Related Work}
\label{sec:relatedWorks}

Influence Maximization Problem (IMP) \cite{kempe2003maximizing} has been extensively studied by many researchers. Given a large-scale and dynamic environment, mining influencers without re-calibration over time becomes a challenging issue. Sun et al. \cite{sun2018multiround} developed a multi-round greedy algorithm which selects seeds by searching all nodes at both global level and within-round level. To be specific, instead of selecting the most influential node at each round, the within-round greedy selection searching a set of seed nodes for each iteration. Yuan et al. \cite{yuan2019efficient} improved the greedy algorithm by using a two-stage framework, in which finding seed candidates and filtering connected nodes before the seed selection. Ding et al. \cite{ding2020influence} raised three greedy algorithms based on a proposed influence cascade model which considers the activation probability of potential influencers. However, greedy algorithm is not scalable, thus, many researchers further developed seeding algorithms to mitigate this issue. Tang et al. \cite{tang2020adiscrete} proposed a discrete shuffled frog-leaping algorithm which is scalable for large-scale networks as it performs a discrete local search at each population. Shang et al. \cite{shang2017CoFIM} sub-modularized influence diffusion process and propose a community-based algorithm which is more efficient than the traditional greedy algorithm. However, these algorithms are not adaptive for dynamic networks. Targeting to the offline events problem, Shi et al. \cite{shi2019location} proposed a location-driven diffusion model and developed the greedy algorithm over integer lattice. Guo et al. \cite{guo2020influence} reduced the average size of the random reverse reachable set generation in IM algorithm which shows a more efficient performance. Rather than computing the influence with the consideration of the network structure, Tian et al. \cite{tian2020deep} proposed a novel deep influence evaluation model to evaluate the user influence. Considering the community-based acceptance probability, Yan et al. \cite{YAN2020116} developed a pipage rounding algorithm with an approximation ratio $(1-1/e)$. Zareie et al. \cite{ZAREIE2020105580} developed an improved cluster rank approach which considers both network structure and users' neighbourhood set. Wen et al. \cite{WEN2020105717} proposed a multi-local dimension method to identify influencers which is based on the fractal property. To evaluate the influence of nodes, Liu et al. \cite{LIU2020105464} proposed a generalized mechanics model with the consideration of both global and local information. Kindled by the competitive  influence maximization problem, Bozorgi et al. \cite{BOZORGI2017149} proposed a community-based algorithm which computes the local influence of each node within its community. Keikha et al. \cite{KEIKHA2020112905} employed deep learning techniques for IMP, they proposed a monotone and submodular algorithm which guarantees an optimal solution. However, most existing scalable seeding algorithms still cannot handle the dynamic structure of a social network. 


By extending the classic influence maximization problem, many studies have been dedicated to adaptive influence maximization problem for dynamic social networks, which aims to address the challenges in regards to the high-speed data transmission and a large population of participants \cite{hafiene2020influential,singer2016influence}. Tong and Wang systematically investigated adaptive influence maximization problem with an emphasis on the algorithmic analysis of the general feedback models \cite{tong2020adaptive}. Tong et al. first formally modelled the dynamic independent IC model and proposed an adaptive hill-climbing strategy \cite{tong2016adaptive}. Han et al. studied adaptive influence maximization problem by developing the first practical algorithms based on a novel AdaptGreedy framework \cite{han2018efficient}. Rodriguez and Sch{\"o}lkopf tackled the influence maximization in temporal networks by proposing INFLUMAX model, which considered the time-series dynamics in the diffusion process \cite{rodriguez2012influence}. However, most of the existing studies focus on the development of adaptive seed selection policies for dynamic networks. Different from the existing research work, we adopt an agent-based evolutionary approach to address the influence maximization problem in a large-scale and dynamic environment, where the solution can be automatically adapted with the evolving of topological structure.

Genetic Algorithm (GA), inspired by the ``survival of the fittest'' theory, is a typical evolutionary computational approach for solving optimization problems which has been applied to IMP in recent years. In GA, an individual which is encoded as a limited size of chromosomes represents a potential seed set while a gene of the chromosome refers to a seed user.  Tsai et al. \cite{tsai2015agenetic} combined GA with the greedy algorithm to improve the effectiveness of IMP solution. In \cite{bucur2016influence}, a simple genetic algorithm was proposed for solving IMP, which shows priors over greedy algorithms as GA obtains more feasible solutions without any assumptions about the network structure. Kromer and Nowakova \cite{kromer2018guided} considered IMP as a fixed-length subset selection problem and developed a new genetic algorithm for IMP, where the network structure has been taken into consideration. Zhang et al. \cite{zhang2017maximizing} proposed a multi-population genetic algorithm which kept the diversity of the solution in LT with a fixed threshold, and the experimental results demonstrated its similar performance to the greedy algorithm. However, almost all the existing GA-based algorithms applied to influence maximization cannot handle the dynamics of social networks. The strategies of narrowing down the researching scope for GA also rely on the global view, i.e., the entire network topological structure. This makes it difficult to handle the large-scale network.

By considering the limitations of the existing seeding algorithms, in this paper, we propose a novel evolutionary approach, which efficiently facilitates the searching process, involving both local views from users and a filtered limited global view, i.e., the influencer pool. The proposed ABEM can mitigate the issues of scalability and topological dynamics. Particularly, we utilize a decentralized approach to narrow down the searching scope of GA, and develop an algorithm to dynamically adapt the problem space according to the changing environment.  

Agent-based Modelling (ABM) is widely acknowledged as an appropriate tool for modelling complex systems by defining the problems at a microscopic level. There are quite a few existing works modelling influence diffusion by using ABM. Li et al. introduced a novel agent-based influence diffusion model and utilize IMP as an application to validate the proposed model \cite{li2016agent}. Based on the same agent-based diffusion model, further research was conducted to exploring how to maintain long-term influences, where ABM assists in capturing both individual's behaviours and influence status at a microscopic level \cite{li2017agentbased}. Li et al. \cite{li2019amultiagent} proposed an enhanced evolution-based backward algorithm for selecting seed users. In this framework, agents share a common knowledge repository and can learn preferences from both shared knowledge and local information. 

Whereas, in almost all the research works mentioned previously, the utilization of ABM aims at modelling the diffusion process, but not for the influencers identification. In other words, ABM is not directly involved in the seeding algorithms. Different from these studies, we leverage ABM as a tool for the influencers ``pre-selection", filtering out noncompetitive candidates. In this sense, ABM can significantly reduce the researching space for GA and address the computational overhead for individual's influential capability evaluation. 

Next, we will give formal definitions required for this research, and formally define the IMP based on the current setting. 

\section{Formal Definitions and Problem Formulation}
\label{sec:preliminaries}
\newcommand*{\field}[1]{\mathbb{#1}}

\subsection{Formal Definitions}
\noindent
\textbf{Definition 1: A dynamic social network} $G_D = \{G(t) | t\in \field{N} \}$ can be modelled as a set of graphs, captured at different time steps. $\forall G(t) \in G_D $,  $G(t)=<V(t),E(t)>$ refers to a directed graph at a time step $t$. The node set $V(t)=\{v_1,v_2,\ldots, v_n\}$ refers to a set of users which are part of the topological structure of $G(t)$ at $t$. Edge set $E(t)$ describes the friendship among these users. Each edge $e_{ij}$ can be represented as a tuple, i.e., $e_{ij}=(v_i,v_j)$, implying a potential influential relationship between $v_i$ and $v_j$. \\

\noindent
\textbf{Definition 2: A user agent} $v_i\in V(t)$ refers to an active, autonomous and interactive entity in dynamic social network $G(t)$. User agent $v_i$ is capable of accessing its local context, incorporating neighbours $\Gamma_{v_i}$ and the relationships with $\Gamma_{v_i}$. In particular, $\Gamma_{v_i}(t)$ describes the adjacent neighbours of $v_i$ at $t$. User agent $v_j \in \Gamma_{v_i}$ if edge $e_{ij} \in E$. Node degree $d_i(t)$ of $v_i$ denotes the size of $v_i$'s neighbourhood $\Gamma_{v_i}(t)$ at $t$. Mathematically, $d_i(t)$ can be represented by using the size of $v_i$'s neighbourhood, i.e., 

\begin{equation}
\label{eq:nodeDegree}
    d_i(t) = |\Gamma_{v_i}(t)|,
\end{equation}

Meanwhile, within a limited local view, user agent $v_i$ conducts influence capability estimation with the assists of neighbourhood, where the influence capability describes the number of users can be influenced by $v_i$. Specifically, given a limited number of influence diffusion level $l$, agent $v_i$ diffuses an influence with a maximum hops of $l$. Agent $v_l$, requiring $l$ hops to reach $v_i$, can provide status feedback to $v_{l-1}$, so on and so forth. Therefore, the influential capability estimated by $v_i$ is denoted as $\sigma(v_i) $.\\

\noindent
\textbf{Definition 3: Influencer pool} $C(t) \subseteq V(t) $ describes a collection of potential influencers at $t$, which is constructed through proactive proposals initiated by user agents $V(t)$. The influencer pool can be recognized as a common knowledge repository, which is accessible by all the user agents. The size of influencer pool varies according to the changing network topological structure, e.g., $|C(t)| $ can be different from $|C(t+1)|$. 

An element $v_c \in C(t)$ describes an influencer candidate. Specifically, user agent $v_c$ proactively estimates the influential capabilities against $\Gamma_{v_c}$ at $t$ and determines whether to propose as one of the potential influencers in $C(t)$. In the current context, user agent $v_c$ intends to nominate itself as a seed candidate only when its degree exceeds a particular degree threshold $\theta_s$ and its influence exceeds a certain percentage, i.e., influence quartile threshold $\theta_q$, of its neighbours $\Gamma_{v_c}(t)$ at $t$. \\

\noindent
\textbf{Definition 4: A seed set} $S_m(t) = \{v_1, v_2, \ldots, v_k \}, S_m(t) \subseteq V(t) $ refers to a set of identified influencers from social network $G(t)$ at $t$ with a limited budget $k$, i.e., $k=|S_m(t)|$. The influencers selection algorithm is named as seeding algorithm. 

In GA, a seed set corresponds to a \textit{``chromosome"} or an \textit{``individual"}. Mapping to the problem space, each chromosome or individual implies a potential solution to the problem. Within a seed set, each element is called a \textit{``gene"}.  \\

\noindent
\textbf{Definition 5: A population} $R_i$ generally refers to a collection of candidate solutions for a pre-defined problem space in GA. In the current setting, a population corresponds to a collection of seed sets, which are recognized as the potential solutions to the influence maximization problem. Population $R_i = \{S_1, S_2, \ldots, S_j \}$ represents the $i-$th generation of the overall evolution process. $R_i$ evolves to the next generation $R_{i+1}$ through the GA operators, and $R_0$ means the initial generation. $g$ refers generation numbers, which indicates the maximum number of evolutions.


\subsection{Problem Description}

Given a dynamic social network $G(t)=<V(t), E(t)>$ and a limited budget $k$, and $t, k \in \field{N}$, the objective is to efficiently select $k$ users as influencers from $G(t)$ at $t$, i.e., $S(t) $, expecting they can spread a pre-defined influence and maximize the impact $\sigma(S(t)) $ across $G(t)$, where $\sigma(S(t)) $ refers to the influence coverage, describing the expected number of influenced users at the end of diffusion process if $S(t)$ is selected as a seed set. 

After $n$ time steps, given $G(t+n) = <V(t+n), E(t+n)>$, the originally identified seed set $S(t)$ needs to be adapted efficiently as $S(t+n)$ to fit the new problem space without re-calculation. Overall, the solution requires to be adapted automatically with the rapid evolution of the online social network.



\section{Agent-Based Evolutionary Model for Mining Influencers}
\label{sec:ABEM}

In this section, we first explain the overall process of mining influencers from social networks by leveraging ABEM. Next, we drill down to the details from both macro-perspective, i.e., user agent's behaviours modelling, and micro-perspective, i.e., adaptive solution optimization, where the algorithms will elaborated as well.  

\subsection{Overall Process of ABEM}

\begin{figure*}[!htb]
	\center
	\includegraphics[width=1.0\textwidth]{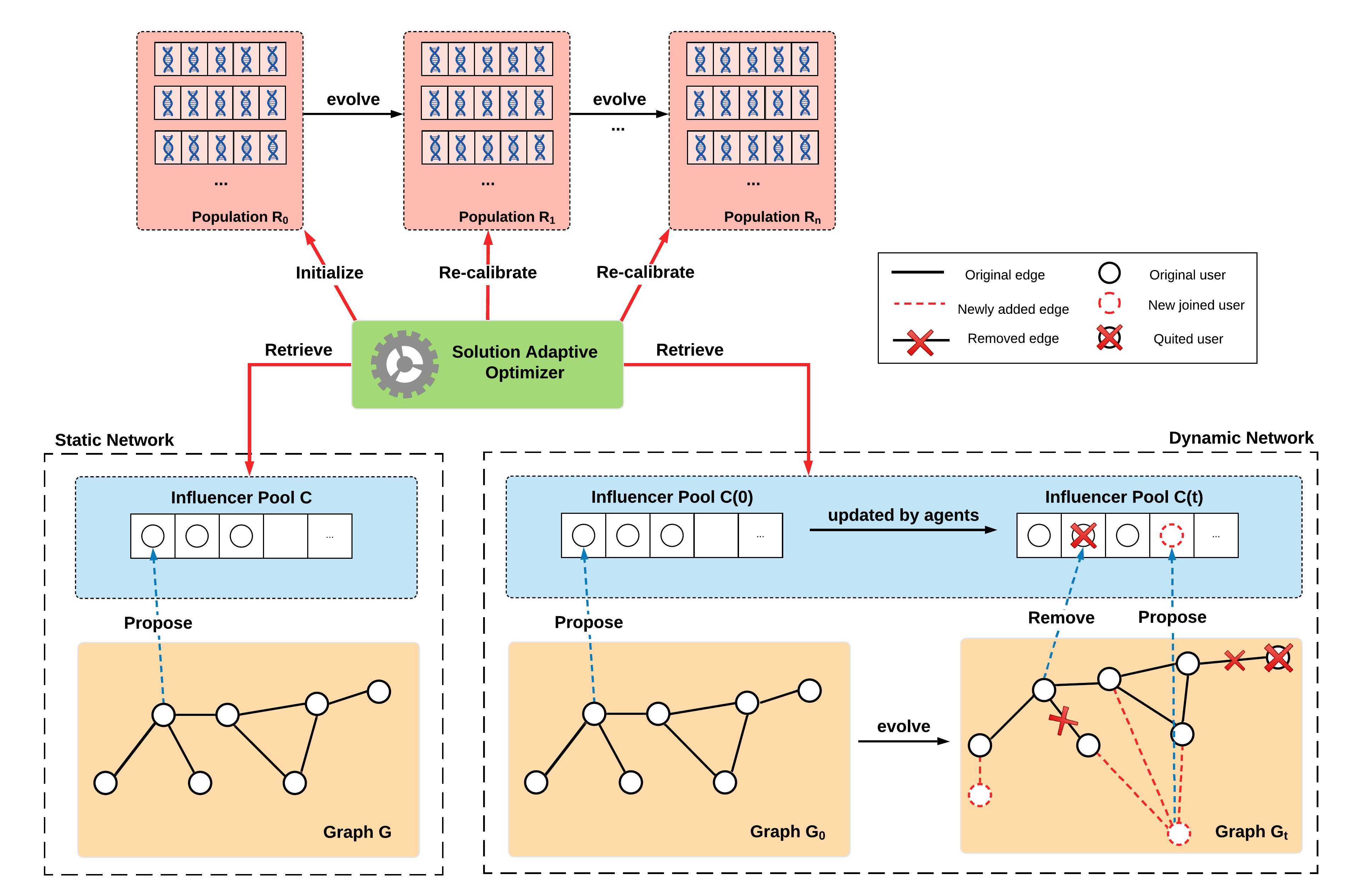}
	\caption{Overall process of ABEM approach}
	\label{fig:ABEMoverallprocess}
\end{figure*}

ABEM facilities the advantages of both agent-based modelling and evolutionary computation, where both agent's local computational power and a centralized optimizer are adopted. Figure \ref{fig:ABEMoverallprocess} illustrates the main idea of ABEM. 

The key process starts from local user agents from Graph $G(0)$, which proactively evaluate its influence capabilities by comparing against the neighbours and decide whether to  propose themselves as influencers via merging into the influencer pool $C(0)$. At the time step $t$, if the local environment of user agent $v_i$ changes, e.g., establishing new links with others, $v_i$ will re-assess the influence capabilities and update the influencer pool $C(t)$ at $t$. In this way, the influencer pool, i.e., the common knowledge repository, always can be kept up-to-date. More importantly, the influencer pool narrows down the searching scope of a large-scale network, leaving only a small amount of data for a centralized component to process.

\begin{table}[!htb]
	\begin{center}
		\caption{Notations Description}
		\label{tab:parameter}
		\begin{tabular}{ccl}
			\toprule
			Notations&Description\\
			\midrule
			$p_a$ & Activation probability of diffusion model \\
			$R_0$ & The initial/first population \\
			$|R_0|$ & The size of $R_0$ \\
			$S_m$ & A chromosome/seed set/solution \\
			$|S_m|$ & The size of chromosome $S_m$ \\
			$p_s$ & Selection rate \\
			$p_c$ & Crossover rate \\
			$p_m$ & Mutation rate \\
			$p_d$ & Degree change rate \\ 
			\bottomrule
		\end{tabular}
	\end{center}
\end{table}

The adaptive solution optimizer plays a pivotal role in ABEM, which is capable of retrieving influencer candidates from the influencer pool in real-time. Meanwhile, it optimizes the solutions generated by GA through re-calibrating the ``genes". With the evolution of GA, the adaptive solution optimizer contributes to the modification of population, leading each generation to fast evolve towards reaching optimal solutions.  


In a nutshell, each user agent undertakes self-evaluation on its influence capability, assisting in identifying the potential influencers. This can effectively handle the dynamics of large-scale social network. Meanwhile, the evolutionary algorithm drives the seed selection process by continuously optimizing the solutions. Next, we elaborate on the modelling of user agents' behaviours and adaptive solution optimization in detail. The notations used in the following algorithms are listed in Table \ref{tab:parameter}. 

\subsection{Influencer Nomination by Autonomous Agents}

In ABEM, users are modelled as autonomous and proactive agents, which are capable of communicating with their neighbours, retrieving information from the local environment and estimating the influential capability. Influencer nomination is initiated by user agents proactively. The outcome of such nomination is revising the influencer pool, keeping it update-to-date with evolving social networks. 

Algorithm \ref{alg:globalSeedCandidate} describes the agent-based influencer nomination process, where all the computations are conducted by agents locally. The inputs include user agent $v_i$, the degree threshold $\theta_s$, and influence quartile threshold $\theta_q$. The output is the updated influencer pool $C(t)$ at time step $t$. Lines 1-2 aim to initialize variables and examine neighbourhood at $t$. Lines 3-6 request neighbours' update-to-date information, including the influential capability, and store the retrieved data to the list. Lines 8-12 calculate influence quartile of $v_i$ by comparing against neighbours and decide how to update the influencer pool. 

\begin{algorithm}[!htb]
	\caption{Agent-based Influencer Nomination Algorithm}
	\label{alg:globalSeedCandidate}
		\begin{flushleft}
			\textbf{Input:} User agent $v_i$, degree threshold $\theta_s $, and  influence quartile threshold $\theta_q$ \\
			\textbf{Output:} Decision of nomination by updating influencer pool $C(t)$ \\
		\end{flushleft}
	
		\begin{algorithmic}[1]
		\State Obtain the up-to-date neighbours $\Gamma_{v_i}$
		\State Initialize a list $L$ for ranking influencers
		\For {$v_j \in \{v_i \cup \Gamma_{v_i} \} $}
		\State Request $v_j$'s influential capability $\sigma(v_j)$
		\State Add $\sigma(v_j)$ to list $L$
		\EndFor
		\State Sort $L$ by $\sigma(v) $ in a descending order
		\If {$(\frac{indexOf(v_i) }{|L|} \geq \theta_q ) \wedge (k_i \geq \theta_s)$}
		\State $C(t)= \{v_i\} \cup C(t)$
		\ElsIf {$v_i \in C(t) $}
		\State $C(t) := C(t) \setminus \{v_i\}  $
		\EndIf
	\end{algorithmic}
\end{algorithm}



\subsection{Adaptive Solution Optimization}

The adaptive solution optimization incorporates two key concurrent processes, i.e., the solution optimization and adaptation. In the former, GA has been adopted to continuously optimize the solutions over time, where the searching space is defined by Adaptive Solution Optimizer (ASO). In the latter, ASO re-calibrates the solutions by considering both the update-to-date influence pool and  the existing outcome. Obviously, ASO plays a critical role in bridging the decisions from user agents and the outputs from the  evolutionary algorithm, which also reflects the main idea of our proposed ABEM for addressing the influence maximization problem.

\begin{algorithm}[!htb]
	\caption{ABEM for Influence Maximization}
	\label{alg:seedSelection}
	\begin{flushleft}
		\textbf{Input:} Dynamic social network $G(t)$, influencer pool $C(t)$, activation probability $p_a$, crossover rate $p_c$, mutation rate $p_m$, and seed set size $k$ \\
		\textbf{Output:}  A seed set $S=\{s_1,s_2, \ldots, s_k\}$.
	\end{flushleft}
	\begin{algorithmic}[1]
		\State Initialize generation $i=0$, Population $R_i$, best solution $S'$
		\State Evaluate individual's influence coverage in $R_0$ using fitness function $\sigma(S_m)$
		\State Find out the best solution from $R_0$ and assign to $S'$
		\While {$!$Termination Condition}
		\State Start re-calibration $Rec(C(t_{now}),R_{i-1}) \to R_i$.
		\State Start selection $Sel(R_i) \to S_m$. 
		\If {random $\xi_c < p_c$}
			\State Start crossover $Cro(S_{fittest},S_m) \to S'_m$.
		\EndIf
		\If {random $\xi_c < p_m$}
			\State Start mutation $Mut(p_m, R_i, C(t_{now})) \to R'_i$, 
		\EndIf
		\State Evaluate influence coverage $\forall S_m \in R_i$, calculate $\sigma(S_m)$.
		\State Find the solution $S_x$ with the greatest influence coverage
		\If {$\sigma(S_x) > \sigma(S')$}
		\State $S' := S_x$
		\EndIf 
		\State $i := i+1$.
		\EndWhile \\
		\Return The fittest individual $S_n \in R_i$.
	\end{algorithmic}
\end{algorithm}   

Algorithm \ref{alg:seedSelection} describes the overall process of mining influencers by facilitating ABEM, where four fundamental operators, including initialization, selection, crossover, and mutation, are tailored to fit the problem space. Furthermore, since the evolution of networks can be reflected by the variation of influencer pool, the solutions will be adapted between two consecutive generations by replacing the outdated candidates with new influencers appeared in the influencer pool, namely, a re-calibration operation.
 
Lines 1-3 initialize a population from the current influencer pool $C(t)$, and evaluate the influence coverage of each individual, where the output of fitness function $\sigma(\cdot)$ represents the estimated influence coverage based on the IC model. Line 4 starts the seeding process, where the termination condition triggered when $\sigma(\cdot)$ of the best solution $S_m$ starts to converge and remains unchanged for a number of generations. Lines 5-12 run through the operators for optimization, where $Sel(\cdot)$, $Cro(\cdot)$, $Mut(\cdot)$ and $Rec(\cdot)$ represent the selection, crossover, mutation and re-calibration operators, respectively. Lines 13-18 evaluate the fitness value of each solution yield by the current generation, and find out the best solution, i.e., a seed set with the greatest influence coverage. Line 20 returns the best solution. 

In the following subsection, we explain the key operators utilized in ABEM, including initialization, selection, crossover, and mutation, as well as re-calibration for solution adaptation.

\subsection{Key Operators}

\noindent
\textbf{Population Initialization.} The initial population of seed sets turns out to be very important since it defines the starting point of exploring the ``best" solution, i.e., a seed set, for the influence maximization problem. ASO generates the initial population $R_0=\{S_1,S_2,\ldots,S_m\}$ with a population size of $|R_0|=m$, i.e., a collection of $m$ candidate solutions (\textit{chromosomes} or \textit{seed sets}). A chromosome $S_m=\{v_1, v_2,\ldots,v_k\}$ corresponds to one seed set, where $k=|S_m|$ denotes seed set size. \\



%


\noindent
\textbf{Selection Operator}. The selection operator assists in identifying a collection of seed set for further improvement, where the fitness value of each individual is considered as the key factor. In other words, seed sets with a higher influence coverage have a greater chance to be selected for the next generation. Moreover, since both original solutions and modified ones via the operators will remain, the number of candidate solutions is much greater than the population size. The selection operator also filters out the ``bad ones" and controls population size. 

Therefore, the selection rate $p_s$ of an individual $S_m$ from population $R_t$ to be selected can be formulated in Equation \ref{eq:selection}. 

\begin{equation}
\label{eq:selection}
p_s(S_m) = \frac{\sigma(S_m)}{\sum\limits^{N}_{i=1}\sigma(S_i)}.
\end{equation}

Algorithm \ref{alg:selectionOperator} describes how the selection operator works. Lines 1-3 check the size variance and determine if the selection continues or not. Lines 4-11 copy over the solutions from an augmented population $R_i$ to $R_i'$ based on the selection probability in Equation \ref{eq:selection}. Lines 12-15 fill $R_i$ if the size of $R_i'$ doesn't reach $|R_i|$.
\begin{algorithm}[!htb]
	\caption{Selection Operator}
	\label{alg:selectionOperator}
	\begin{flushleft}
		\textbf{Input:} An augmented population $R_i$ and target population size $|R_i'|$.\\
		\textbf{Output:} Population $R_i'$ after selection. 
	\end{flushleft}
	\begin{algorithmic}[1]
		\If {$|R_i| == |R_i'|$}
		\State return $R_i$
		\EndIf
		\State Initialize $R_i' := \emptyset$
		\For{$\forall S_m \in R_i$}
		\State Estimate the influence capability of $R_i(n)$, i.e., $\sigma(R_i(n))$.
		\State Calculate selection rate $p_s(S_m)$ using Equation \ref{eq:selection}
		\If{Random decimal $\xi_s < p_s(S_m)$}
		\State $R_i' := R_i' \cup \{S_m\}$
		\EndIf
		\EndFor
		\While {$|R_i'| < |R_i|$}
		\State Select $S_x, \sigma(S_x) \geq \forall S_i \in R_i \wedge S_x \notin R_i'$ 
		\State $R_i' := R_i' \cup \{S_x\}$
		\EndWhile
		\State \textbf{Return} $R_i'$. 
	\end{algorithmic}
\end{algorithm}

\noindent
\textbf{Crossover Operator.} The crossover operation in the influence maximization problem recombines two seed sets (parents) and generates two new solutions (offspring). In other words, two selected seed sets exchange the influencers at a random slicing point and produce two new seed sets. Mixing two solutions may cause duplicated elements in a seed set. Thus, a repair function is required to fix the solution by replacing the duplicated influencer with a random user from the influencer pool. 

\begin{algorithm}[!htb]
	\caption{Crossover Operator}
	\label{alg:crossoverOperator}
	\begin{flushleft}
		\textbf{Input:} Population $R_i$, influencer pool $C(t)$, and crossover rate $p_c$.\\
		\textbf{Output:} Population $R'_i$.
	\end{flushleft}	
	\begin{algorithmic}[1]
		\For{$\forall S_m \in R_i$}
		\If{Random decimal $\xi_c > p_c$}
		\State next
		\EndIf
		\State Select a seed set $S_c \in R_i \setminus \{S_m\}$ which is the best solution in $R_i$
		\State Randomly select a slicing point $\xi_s$
		\State Initialize two offspring solutions, i.e., offspring$_1$, offspring$_2$
		\For {$i$ in range(0,$\xi_s$)}
		\State offspring$_1$.add($S_m$.get($i$))
		\State offspring$_2$.add($S_c$.get($i$))
		\EndFor
		\For {$i$ in range($\xi_s, |S_m|$)}
		\State offspring$_1$.add($S_c$.get($i$))
		\State offspring$_2$.add($S_m$.get($i$))
		\EndFor
		\While{$|$offspring$_1|<|S_m|$}
		\State var newgene := $C_t$.get(random($|C_t|$))
		\State offspring$_1$.add(newgene)
		\EndWhile
		\While{$|$offspring$_2|<|S_m|$}
		\State var newgene := $C_t$.get(random($|C_t|$))
		\State offspring$_2$.add(newgene)
		\EndWhile
		\State $R_i := R_i \cup \{$offspring$_1$, offspring$_2\}$
		\EndFor
		\\
		\Return $R_i$.
	\end{algorithmic}
\end{algorithm}

Crossover operator is described in Algorithm \ref{alg:crossoverOperator}. Lines 2-3 check if the current seed set $S_m$ is selected for crossover. Lines 5-7 prepare the operation by obtaining another seed set, generating a slicing point and initialize two offspring. Lines 8-15 conduct crossover. Lines 16-23 repair the generated seed set by adding users from the influencer pool. Because offspring is modelled as a hash set where duplicated items remain a single copy. In other words, the offspring with a lower size requires to be ``fixed". Line 24 expands the current generation by appending the newly generated offspring and Line 26 returns the updated $R_i$ as $R_i'$. \\

\noindent
\textbf{Mutation Operator.} The mutation operator works on an individual user of a seed set, replacing a specific seed (user) with another potential influencer. This operator helps to maintain the diversity of seed sets from one generation to the next, which enables ABEM to have a wide range of feasible solutions, avoiding rapid coverage to a local optimal solution. Specifically, a seed is supposed to be substituted by a randomly selected seed candidate from the current influencer pool with a certain probability. The operator is described in Algorithm \ref{alg:mutationOperator}. \\



\begin{algorithm}
	\caption{Mutation Operator}
	\label{alg:mutationOperator}
	\begin{flushleft}
		\textbf{Input:} A population $R_i$, influencer pool $C(t)$, and mutation rate $p_m$\\
		\textbf{Output:} A mutated population $R'_i$.
	\end{flushleft}	
	\begin{algorithmic}[1]
		\For{$S_m \in R_i$}
		\For{$v \in S_m$}
		\If{Random decimal $\xi_m < p_m$}
		\State Randomly select $c \in C(t) \wedge c \notin S_m $.
		\State $c \to v$
		\EndIf
		\EndFor
		\EndFor\\
		\Return $R'_i$.
	\end{algorithmic}
\end{algorithm}

\noindent
\textbf{Re-calibration Operator}. The re-calibration operator aims to adapt the existing population based on the changing environment. As the influential capabilities of the seeds introduced to the current population vary over time, it is essential to update the existing solution by replacing partial ``out-dated" influencers. Specifically, the re-calibration operator checks through all the seed sets within a population, figuring out the users whose influential capabilities are degraded significantly. Such users will be replaced with those who are newly introduced to the influence pool. 

Algorithm \ref{alg:re-calibrationOperator} describes the re-calibration process. The inputs include user set $V(t)$ at time step $t$, influencer pool $C(t)$, and the current population $R_i$ to be adapted. The output is re-calibrated population $R_i'$. Lines 3-5 identify the users from $R_i$, who have quit the network, and replace these users with a randomly selected user from the influence pool. Lines 6-13 initiate adaptation based on the estimated degree variation rate.

\begin{algorithm}
	\caption{Re-calibration operator}
	\label{alg:re-calibrationOperator}
	\begin{flushleft}
		\textbf{Input:} A user set $V(t)$, influencer pool $C(t)$, current population $R_{i}$\\
		\textbf{Output:} Re-calibrated population $R'_{i}$ 
	\end{flushleft}
	\begin{algorithmic}[1]
		\For{$\forall S_m \in R_i$}
			\For{$\forall v \in S_m$}
				\If{$\nexists v \wedge v\in V(t)$}
					\State Randomly select $c \in C(t) \wedge c \notin S_m $
					\State $c \to v$
				\ElsIf {$v \notin C(t)$}
				\State Generate a random decimal $\xi_d$
				\State Calculate degree change rate $p_d = 1- \frac{|v|}{|v'|}$
					\If {$\xi_d < p_d$}
					\State Randomly select $c \in C(t) \wedge c \notin S_m $
					\State $c \to v$
					\EndIf
				\EndIf
			\EndFor
		\EndFor
		\State \textbf{Return} $R'_i$.
	\end{algorithmic}
\end{algorithm}

\section{Experiments} \label{sec:experiments}

In this section, four experiments are conducted to evaluate the performance of ABEM. The first experiment analyzes the convergence of ABEM with different experimental settings. In the second experiment, we compare the performance of ABEM against several baselines in static networks. The third experiment further explores the parameter settings of ABEM. The last experiment aims to evaluate the performance of ABEM in a dynamic environment. The following subsections introduce the experimental settings, presents the experiment details and discuss the results, respectively.

\subsection{Experimental Settings}
\label{sec:exp_settings}

Three real-world datasets are utilized for the experiments, including Ego-Facebook \cite{panzarasa2009patterns}, Git \cite{rozemberczki2019multiscale}, and Flixster \cite{jamali2010amatrix}. The properties of these datasets are described in Table \ref{tab:networks}, and the parameters of ABEM are listed in Table \ref{tab:ABEM_parameter}. 
\begin{table*}[!htb]
	\begin{center}
		\caption{Datasets}
		\label{tab:networks}
		\begin{tabular}{clll}
			\toprule
			Network&No. of nodes&No. of edges&Type\\ 
			\midrule
			Ego-Facebook \footnote{https://snap.stanford.edu/data/ego-Facebook.html}& 1,899 & 20,296 & static, undirected\\
			Git\footnote{http://snap.stanford.edu/data/github-social.html}&13,419&59,259& static, undirected\\
			Flixster \footnote{http://www.flixster.com/}& 14,231 & 79,265 & dynamic, directed\\
			\bottomrule
		\end{tabular}
	\end{center} 
\end{table*}

\begin{table*}
	\begin{center}
		\caption{Parameters of ABEM}
		\label{tab:ABEM_parameter}
		\begin{tabular}{cll}
			\toprule
			Parameter&Description&Value\\ 
			\midrule
			$p_c$ & Crossover rate & 1\\
			$p_m$ & Mutation rate & 0.1\\
			$|R_t|$ & Population size, the number of chromosomes in one population.& 50\\
			$g$ & Generation numbers, the maximum number of iterations & 1000\\
			\bottomrule
		\end{tabular}
	\end{center} 
\end{table*}
Influence coverage, i.e., the classic evaluation metrics of the influence maximization problem, is adopted for all the experiments. Influence coverage refers to the number of users activated (influenced) by the identified influencers. On top of that, we use elapsed running time as evaluation metrics in Experiment 3, indicating the time cost for the algorithm to find the solutions. To conclude, influence coverage and running time represent the effectiveness and efficiency of the proposed algorithm, respectively.

The following baselines for the influence maximization problem are utilized for performance comparison, where the greedy algorithm is recognized as one of the strongest baseline. 
\begin{itemize}
	\item Greedy: Each seed is selected by iterating all the users, aiming at reaching the maximum influence marginal gain. The greedy algorithm is not scalable as it relies on enormous times of Monte-Carlo simulations.
	\item Degree-based selection: Users with the highest degree will be selected as influencers. 
	\item Degree Discount Heuristics (DDH): The seeds are selected by deterministic degree strategy. This algorithm is developed based on the idea that users with high degree normally cluster together. Hence, it is not necessary to select all of them  \cite{chen2009efficient}.
	\item Genetic Algorithm (GA): The traditional GA without any optimization or tailoring. Specifically, the seeds are selected after a number of evolution with classic GA operators. 
	\item GA with influence pool: The traditional GA with an optimized initial population, where the solutions are initialized based on the influencer pool. 
	\item Random: The seeds are randomly selected at each time step. The executing time is fastest but with a normally lowest influence coverage as it is not based on any heuristics.
\end{itemize}

\subsection{Experiment 1: Convergence analysis}
\label{sec:exp1}
\begin{figure*}[!htb]
	\centering
	\subfigure[ABEM in Git]{\includegraphics[width=0.32\textwidth]{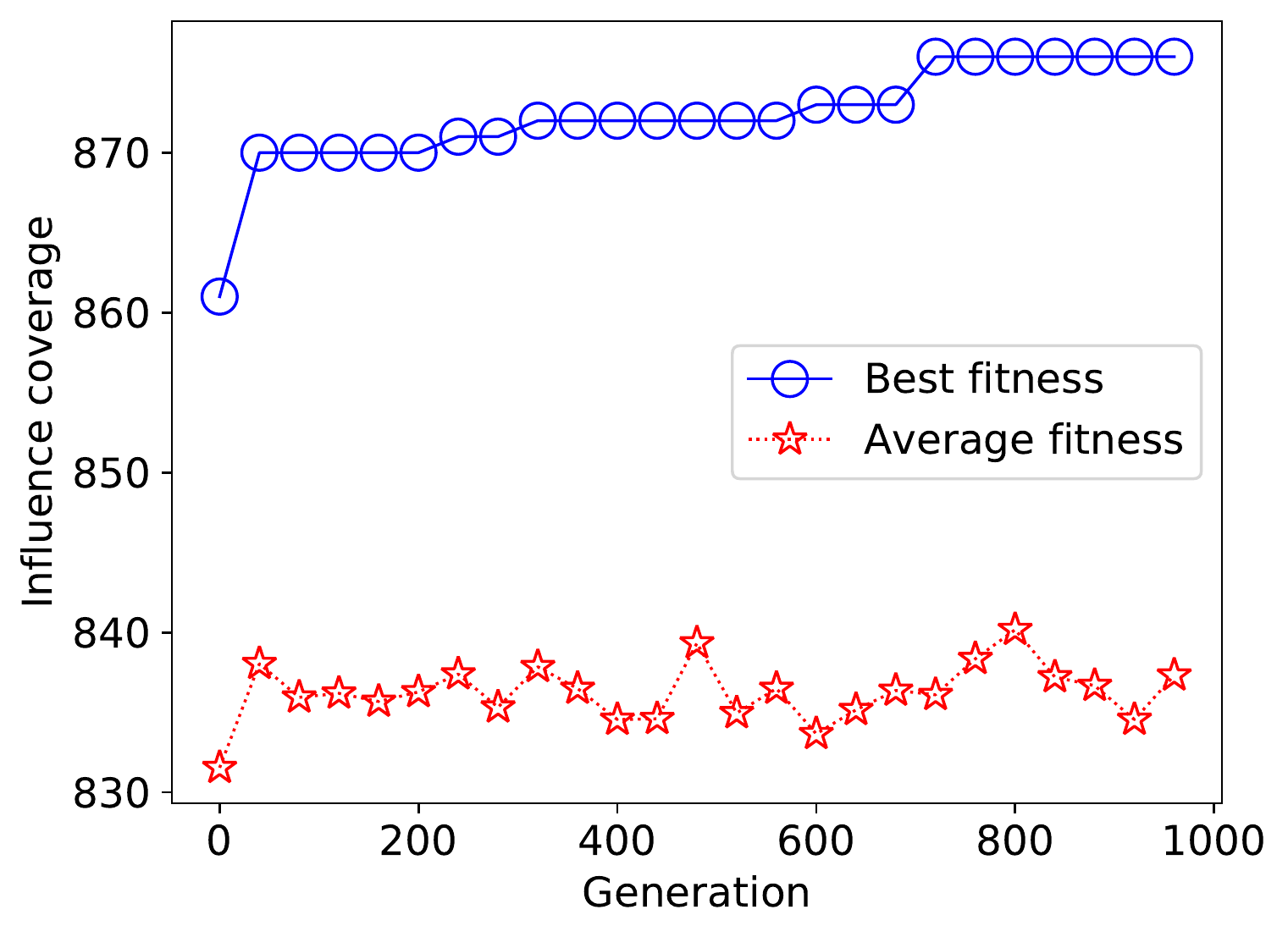}\label{fig:exp1gitabem}}
	\subfigure[GA in Git]{\includegraphics[width=0.32\textwidth]{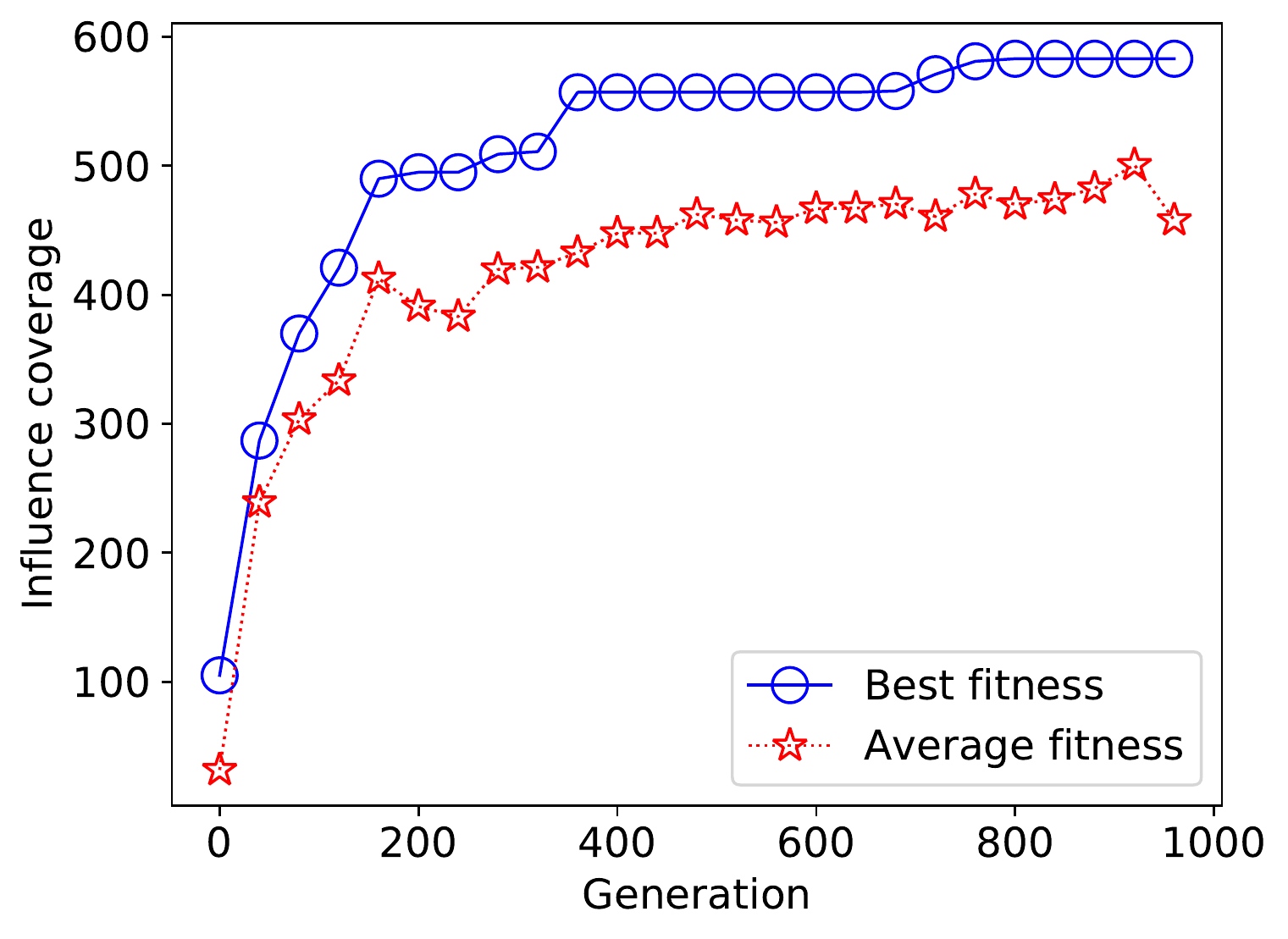}\label{fig:exp1gitga}}	
	\subfigure[GA with influence pool in Git]{\includegraphics[width=0.32\textwidth]{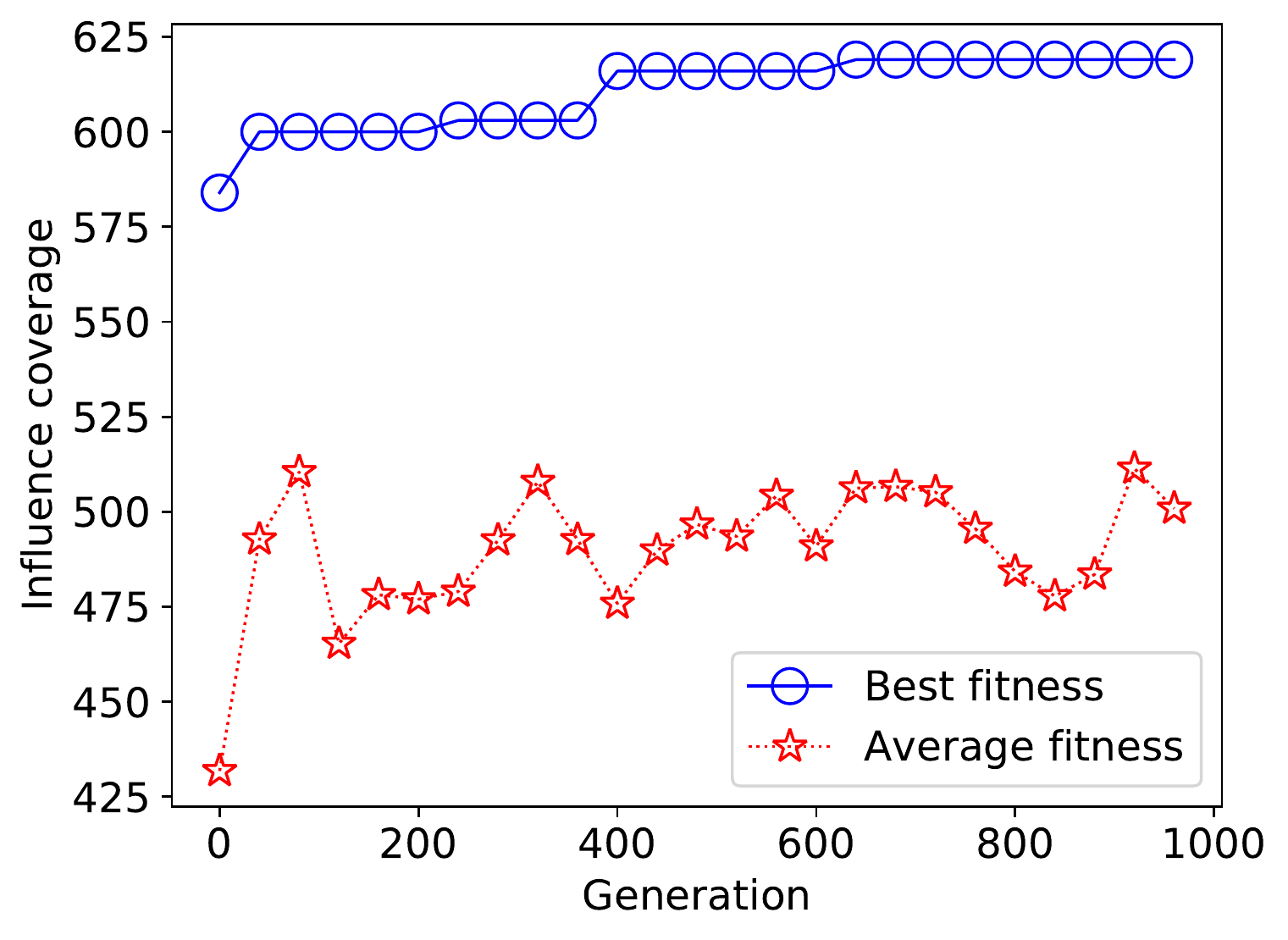}\label{fig:exp1gitgapool}}\\
	\subfigure[ABEM in Ego-Facebook]{\includegraphics[width=0.32\textwidth]{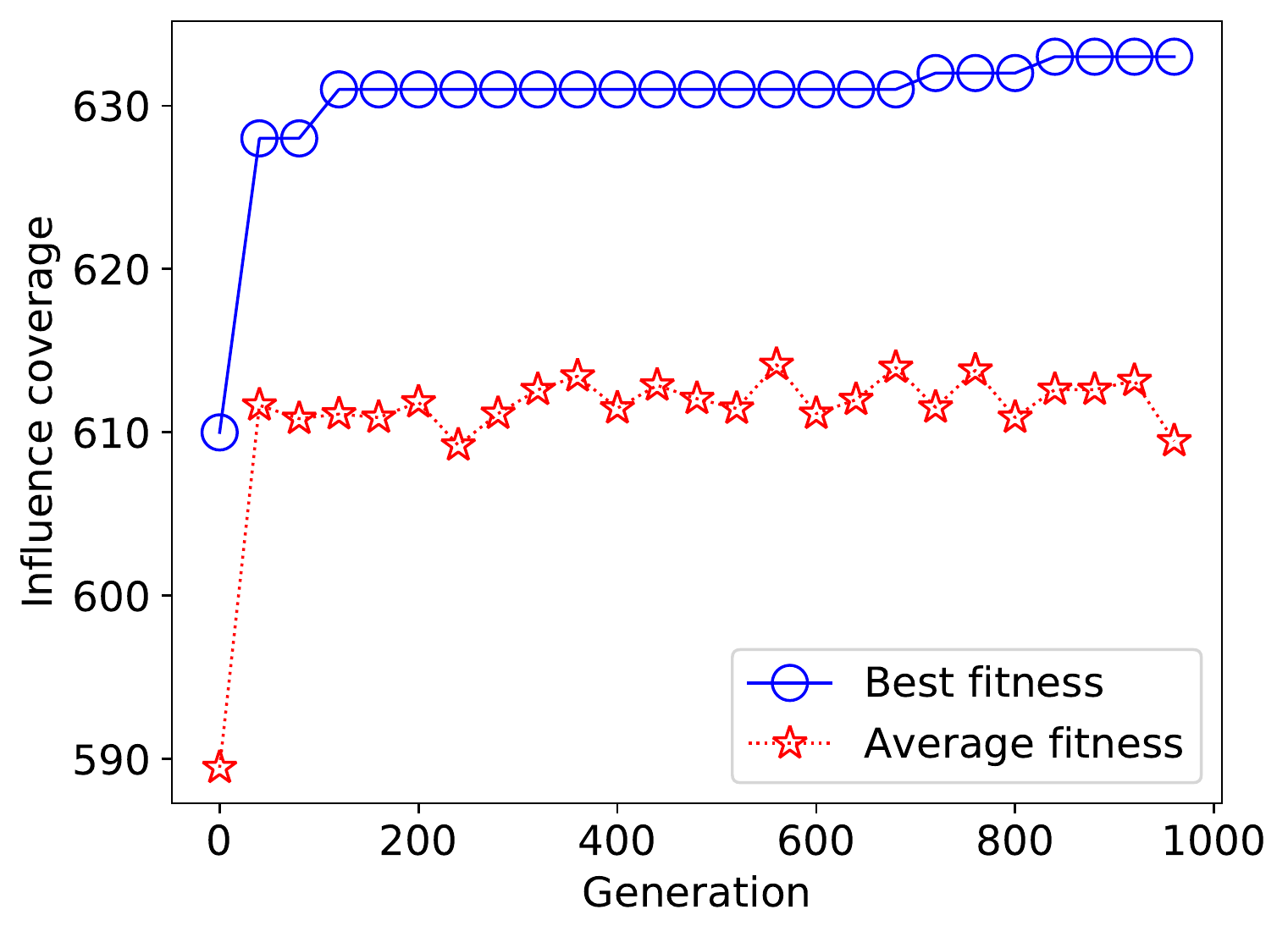}\label{fig:exp1fbabem}}
	\subfigure[GA in Ego-Facebook]{\includegraphics[width=0.32\textwidth]{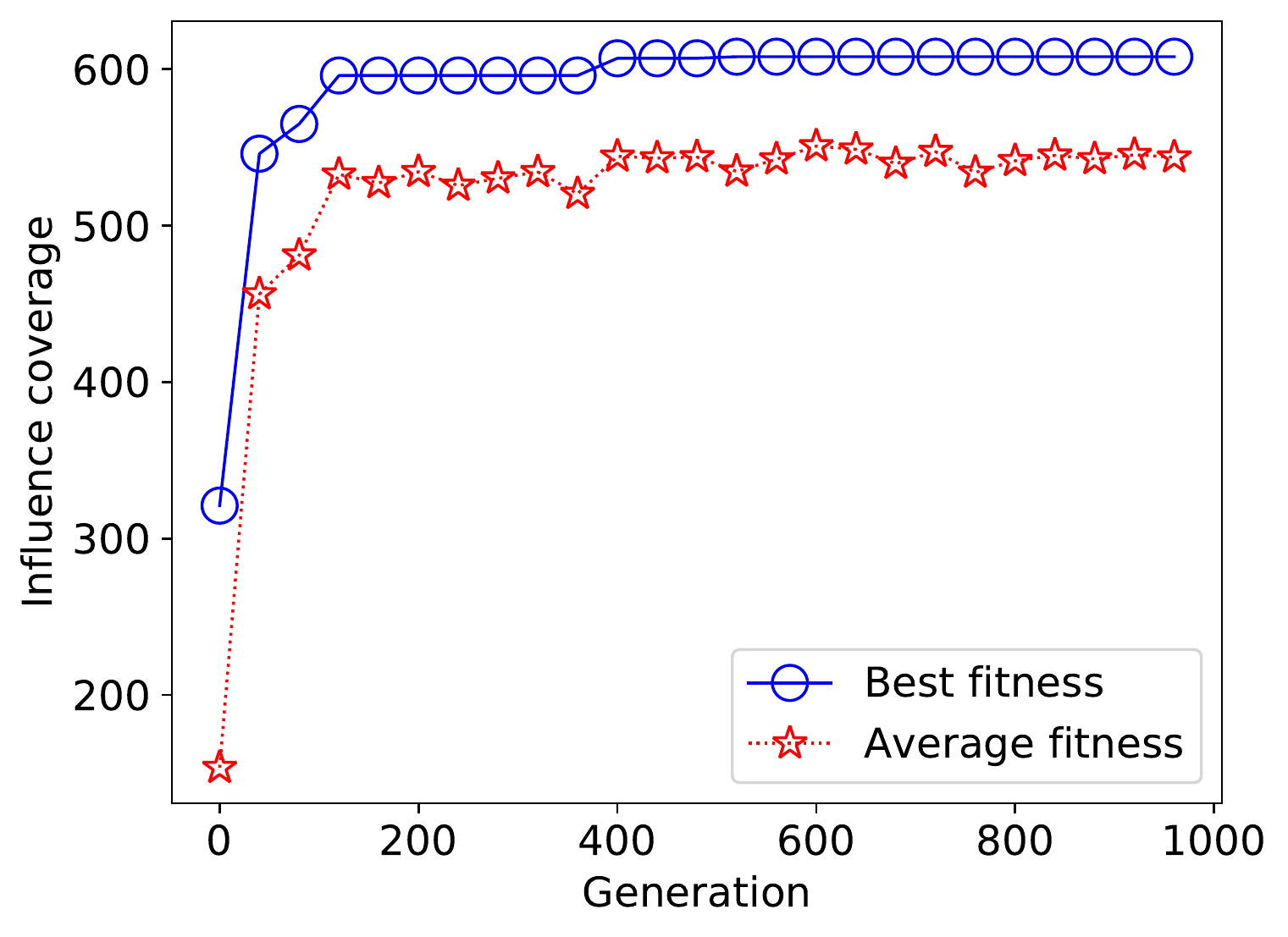}\label{fig:exp1fbga}}
	\subfigure[GA with influence pool in Ego-Facebook]{\includegraphics[width=0.32\textwidth]{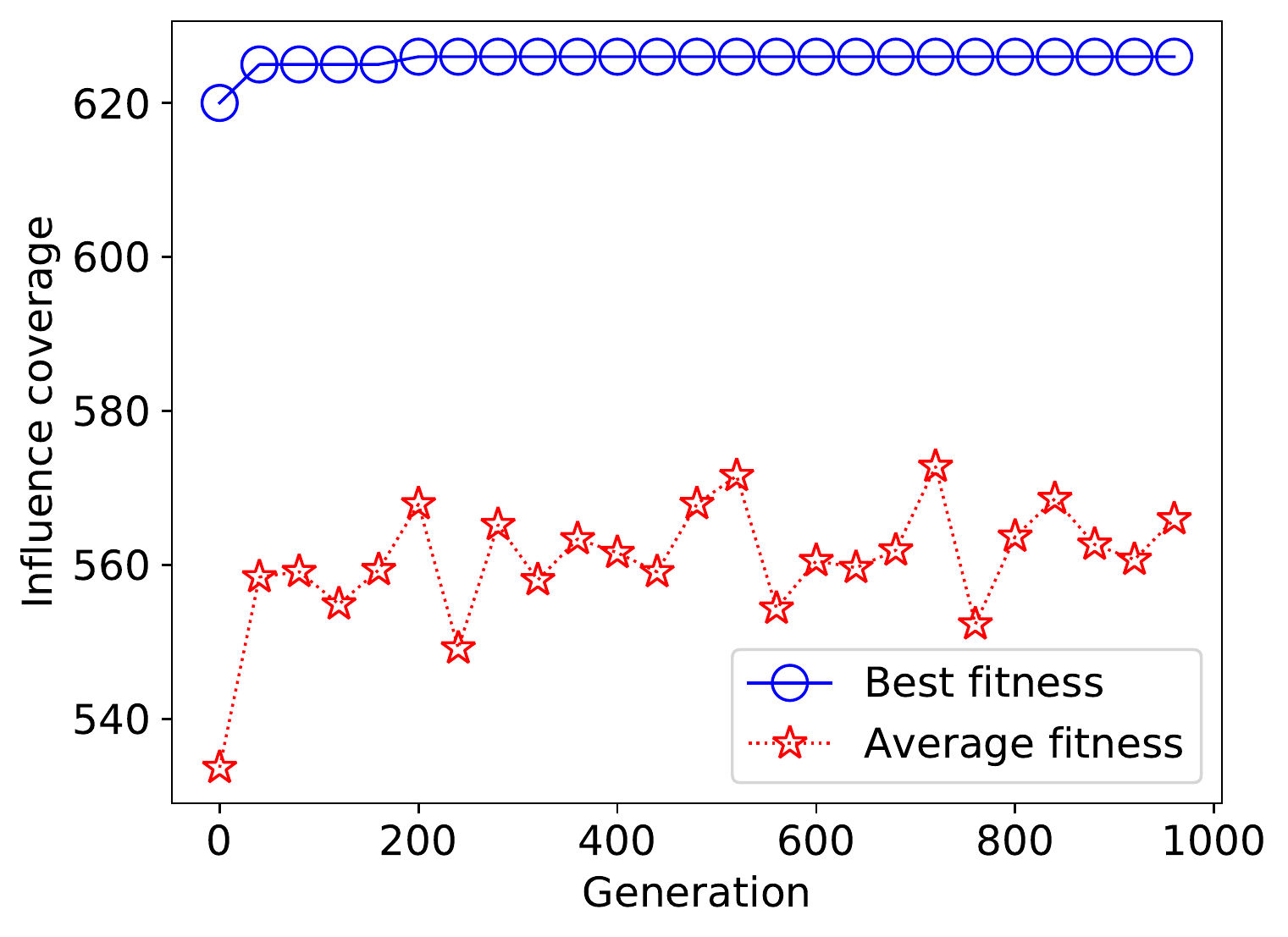}\label{fig:exp1fbgapool}}\\
	\caption{Convergence analysis}
	\label{fig:convergence_Git}
\end{figure*}
Experiment 1 analyses the convergence of ABEM by tracking the evolution pattern of the continuously optimized solutions, i.e., seed sets. For each generation, both average influence coverage (average fitness) and the highest influence coverage (the best fitness) are estimated. The evolutionary algorithms are validated by using two datasets, i.e., Git and Ego-Facebook. The experiment also defines a fixed number of generations, i.e., 1000.


Figure \ref{fig:convergence_Git} shows the evolutionary trends for each population by facilitating ABEM, GA and GA with influencer pool. It is evident that after a series of generations, all the three algorithms start to converge, and finally reach an optimal solution. Furthermore, ABEM demonstrates the best performance of all, which can be revealed by comparing the influence coverage of the best solution against others at the end of 1000 generations. 

The experimental results also implicitly reveal the advantages of ABEM. First, ABEM converges much faster than that of GA, which can be observed by comparing Figure \ref{fig:exp1gitabem} with Figure \ref{fig:exp1gitga}, or Figure \ref{fig:exp1fbabem} with Figure \ref{fig:exp1fbga}. The reason is ABEM leverages the influencer pool for initialization, which enables ABEM to start with a higher point and have a better chance to obtain an optimal solution at speed. Second, ABEM still has a greater chance to improve the existing solutions even reaching a convergence status. Whereas, GA with influencer pool almost makes no chances after converge. This is because the ABEM clearly defines a scope of searching influencers, but the other algorithms conduct the searching process in the world. Therefore, in ABEM, the average fitness of all the populations always appear higher than those of the others. Third, ABEM demonstrates greater computational efficiency. Based on the oscillation degree of average fitness, ABEM shows a relatively stable trend. Whereas, the average fitness of other algorithms fluctuates significantly. The reason behind also relies on the searching scope. In ABEM, the influencer exploration scope is narrowed down by the individual user agents, which greatly reduces the centralized computational cost. By contrast, the other algorithms have to handle a larger scope with dramatic changes in the population.


\subsection{Experiment 2: ABEM valuation in static networks}
\label{sec:exp2}
Experiment 2 aims to evaluate the performance of ABEM in static online social networks. We compare ABEM against the baselines introduced in Subsection \ref{sec:exp_settings}, where two static networks, i.e., Ego-Facebook and Git, are adopted. In this experiment, the seed set size $k$ ranges from 5 to 10 with a step of 5. On top of that, it facilitates the IC model as the influence diffusion model, with a uniform probability of 0.1. 

\begin{figure}[!htb]
	\centering
	\subfigure[Ego-Facebook]{\includegraphics[width=0.48\columnwidth]{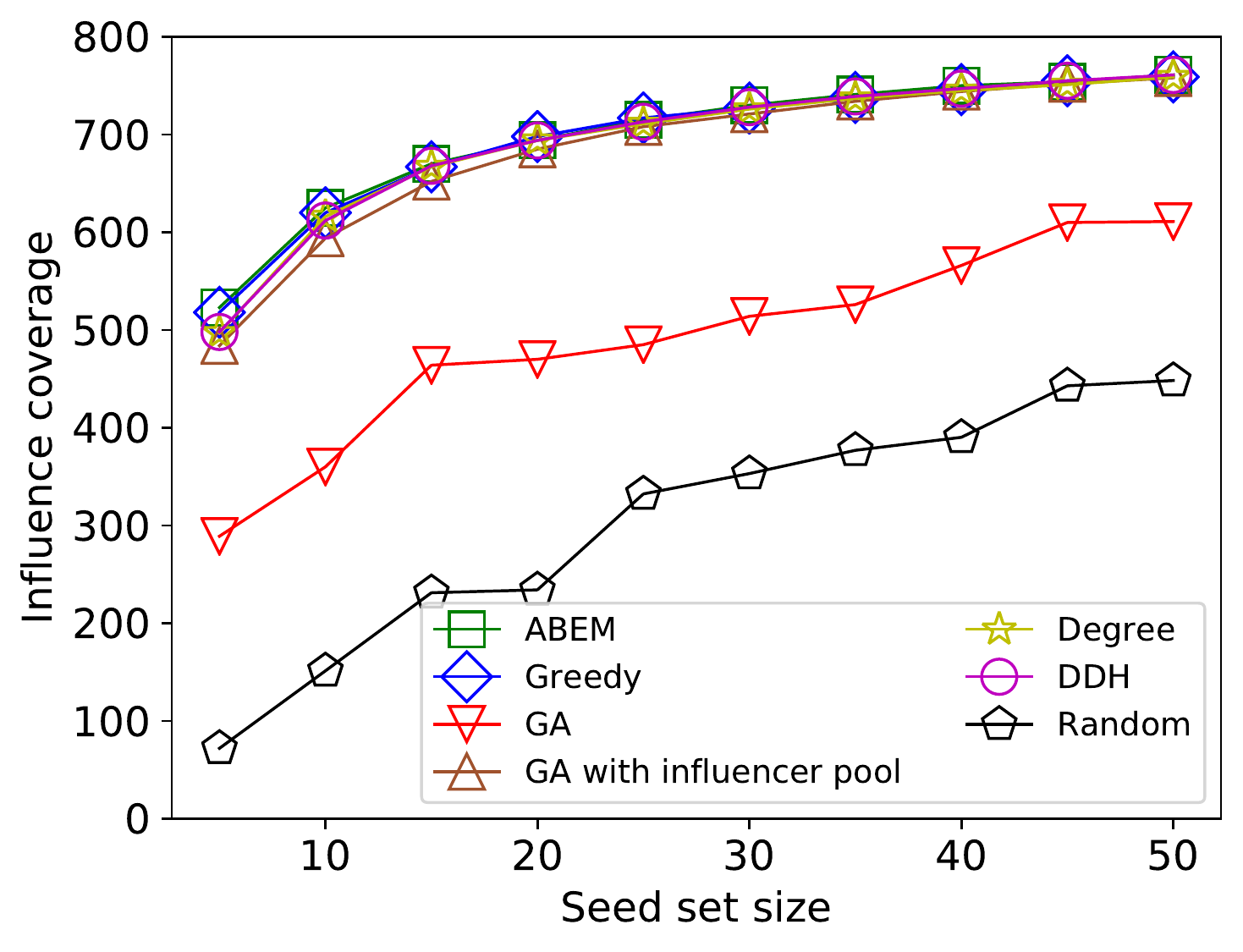}\label{fig:exp2_fb}}
	\subfigure[Git]{\includegraphics[width=0.48\columnwidth]{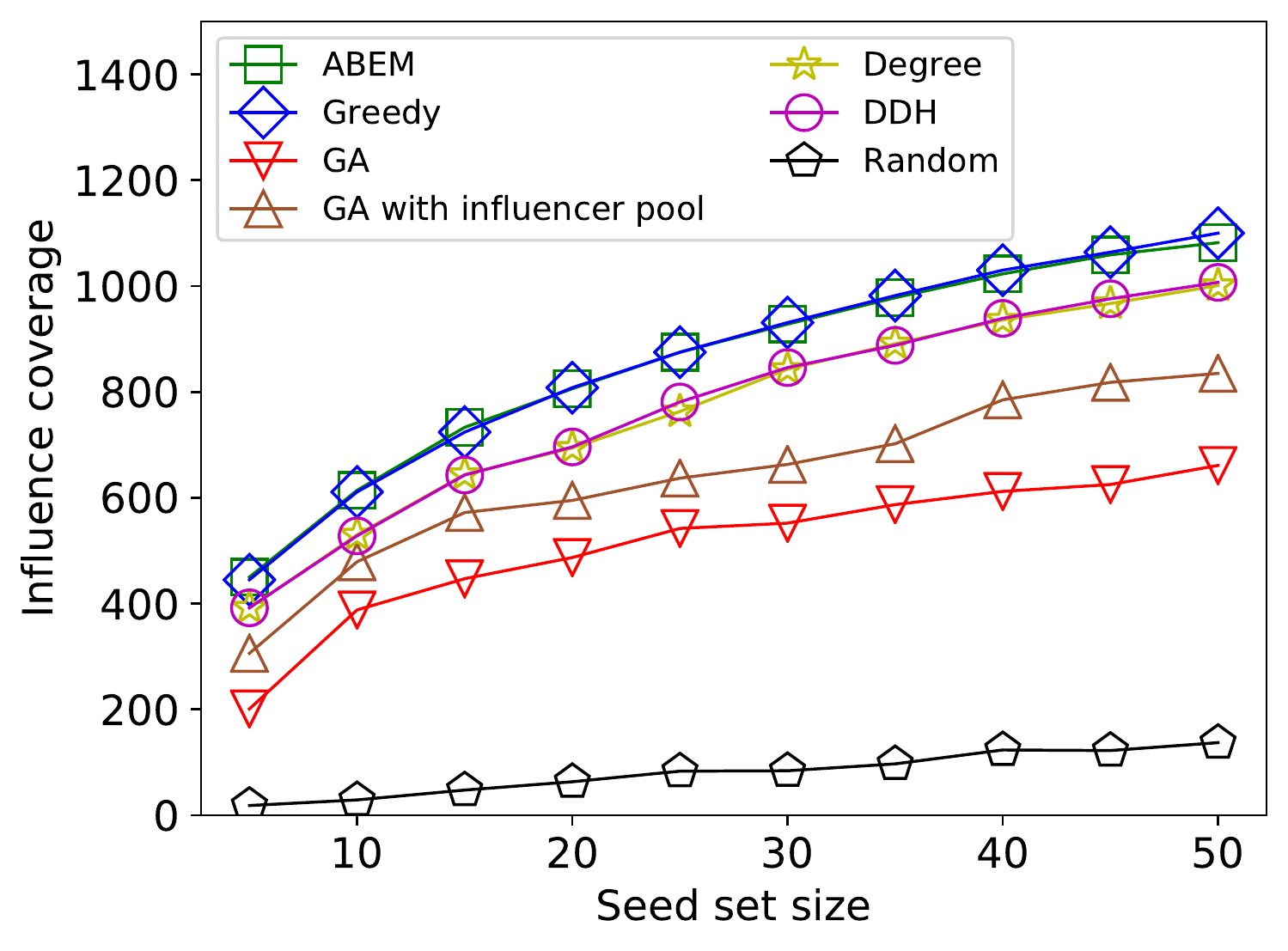}\label{fig:git}}\\
	\caption{Influence maximization with different seed set size}
	\label{fig:exp2}
\end{figure}

The experimental results on Ego-Facebook and Git are demonstrated in Figures \ref{fig:exp2_fb} and \ref{fig:git}, respectively. As aforementioned, the greedy selection has been recognized as one of the strongest baselines in the influence maximization problem, but not scalable. As can be seen from both figures that the greedy selection yields the best performance of all in two datasets. Despite carrying out a similar performance as greedy, ABEM is capable of mitigating the scalability issue since the major computations are distributed to the individual agents and the searching scope is limited. 

In Figure \ref{fig:exp2_fb}, no significant performance difference can be observed among ABEM, Degree, DDH, and Greedy selection. This is because the size of the Ego-Facebook network is small, and the identified seed sets are also similar. Despite this, ABEM performs slightly better than degree and DDH. Give a relatively larger network, in Figure \ref{fig:git}, ABEM outperforms other baselines and demonstrates a similar performance as the greedy selection. Notably, when a small seed set is required, e.g., $k=\{5,10,15\}$, the performance of ABEM can exceed the greedy selection.

\subsection{Experiment 3: Parameter analysis}
\label{sec:exp3}
In Experiment 3, we further investigate the performance of ABEM by varying the parameters, including the number of generation $g$, the degree threshold $\theta_s$, and influence quartile threshold $\theta_q$. 

\begin{figure}[!htb]
	\centering
	\subfigure[Ego-Facebook]{\includegraphics[width=0.48\columnwidth]{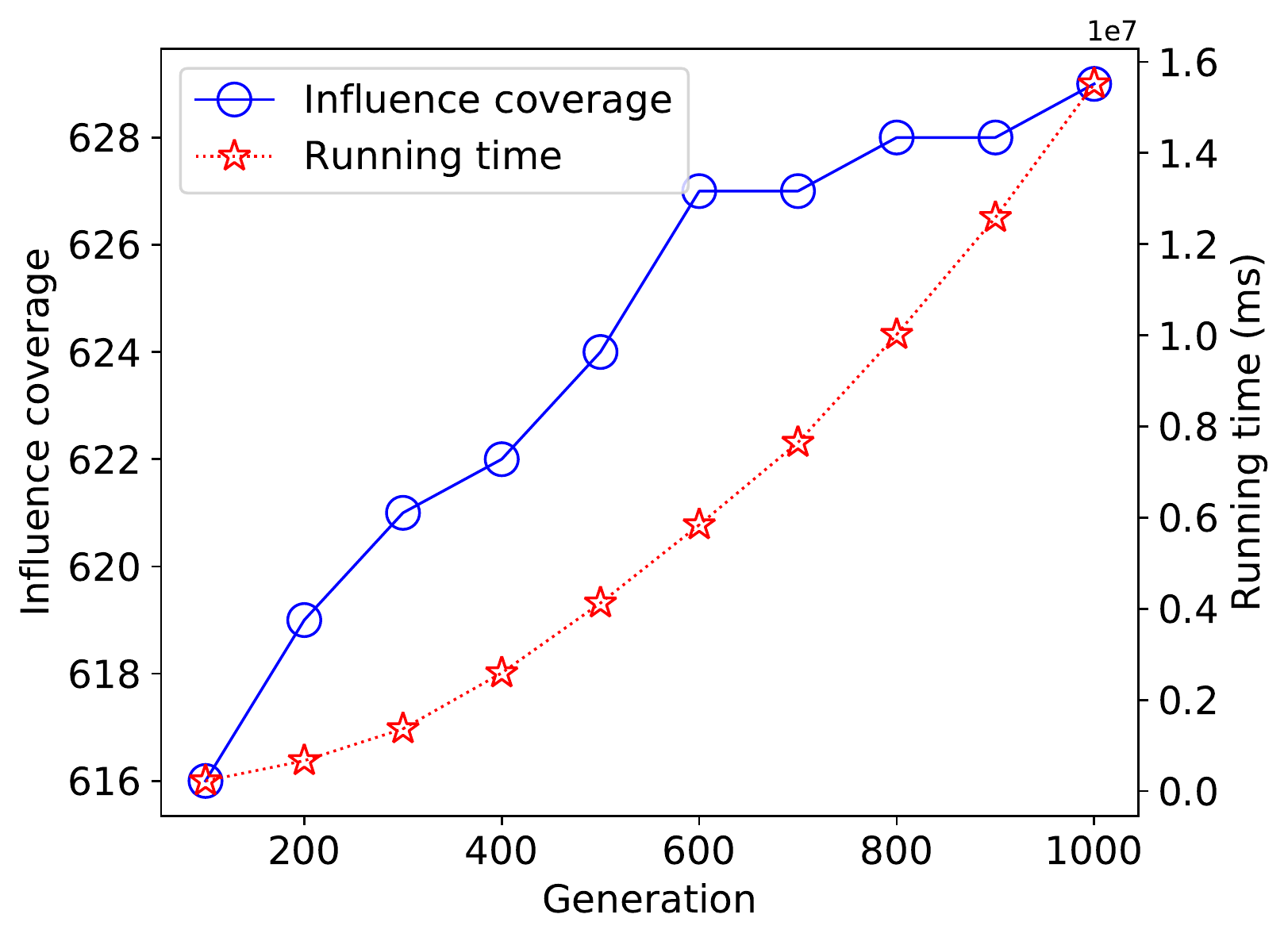}}	
	\subfigure[Git]{\includegraphics[width=0.48\columnwidth]{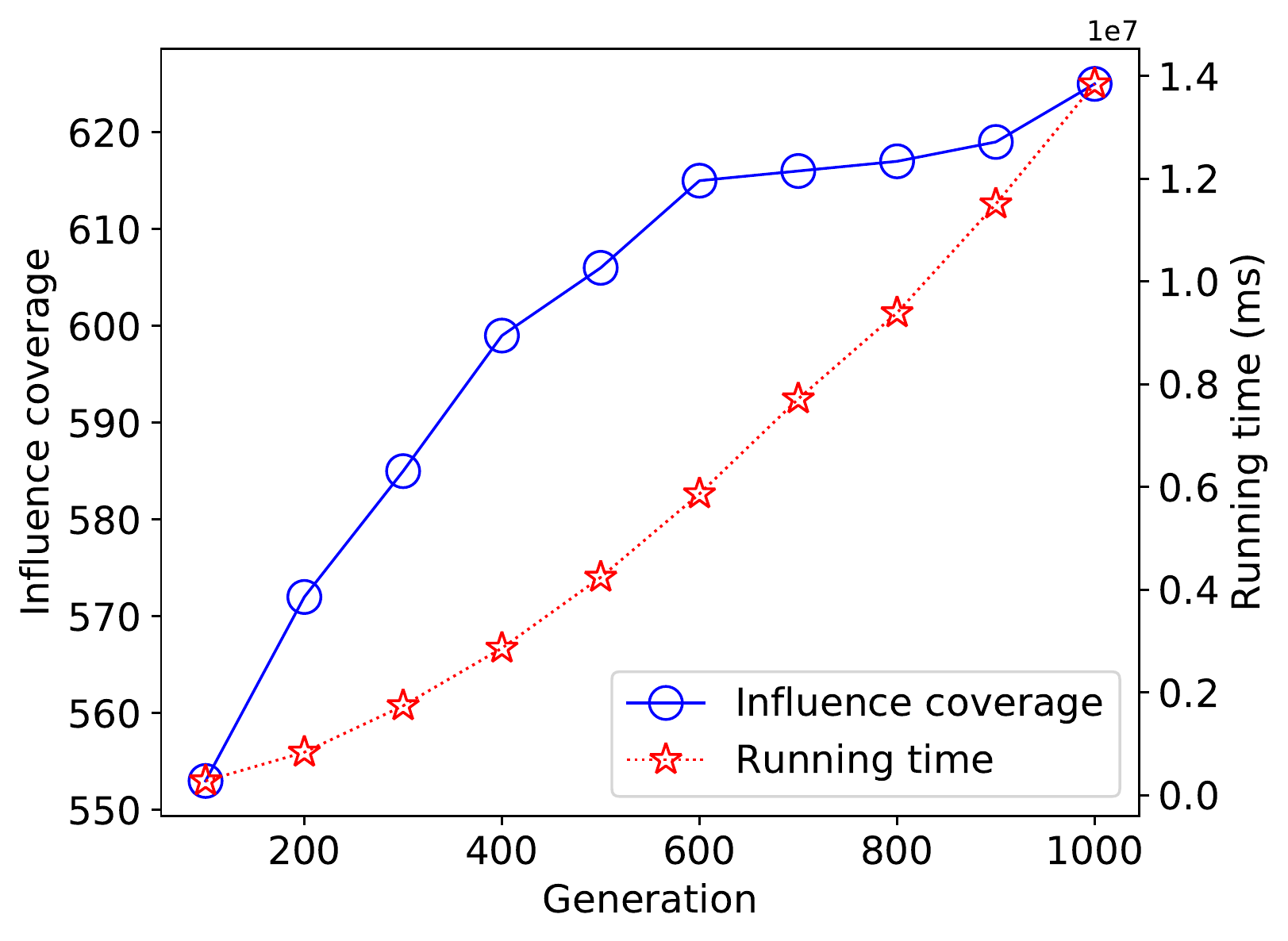}}\\
	\caption{ABEM performance with different generation numbers}
	\label{fig:genNum_git}
\end{figure}

First, we analyze the impact on the influence coverage and running time by increasing the number of generations. As we can observe from Figure \ref{fig:genNum_git} that with the rise of evolutionary generation, the elapsed running time increases linearly. The influence coverage also demonstrates a steady escalation trend, with slight improvement after 500 generations. This is due to the fact that almost all the potential influencers are incorporated into the population, and it takes time to figure out a better seed set by re-organizing the existing influencers. 

Second, we deeply investigate the influencer pool size variations by adjusting the degree threshold $\theta_s$ and influence quartile threshold $\theta_q$. Both parameters control the individual's ``propose as an influencer" behaviour, which directly impacts the size of influencer pool. Subsequently, it determines the searching scope and influences the performance of ABEM. A high degree threshold $\theta_s$ implies only those with large neighbour size can be proposed as an influencer. Likewise, a high quartile threshold $\theta_q$ allows the users who are influencers in their social circle to join the influencer pool. 

\begin{figure}[!htb]
	\centering
	\subfigure[Ego-Facebook]{\includegraphics[width=0.48\columnwidth]{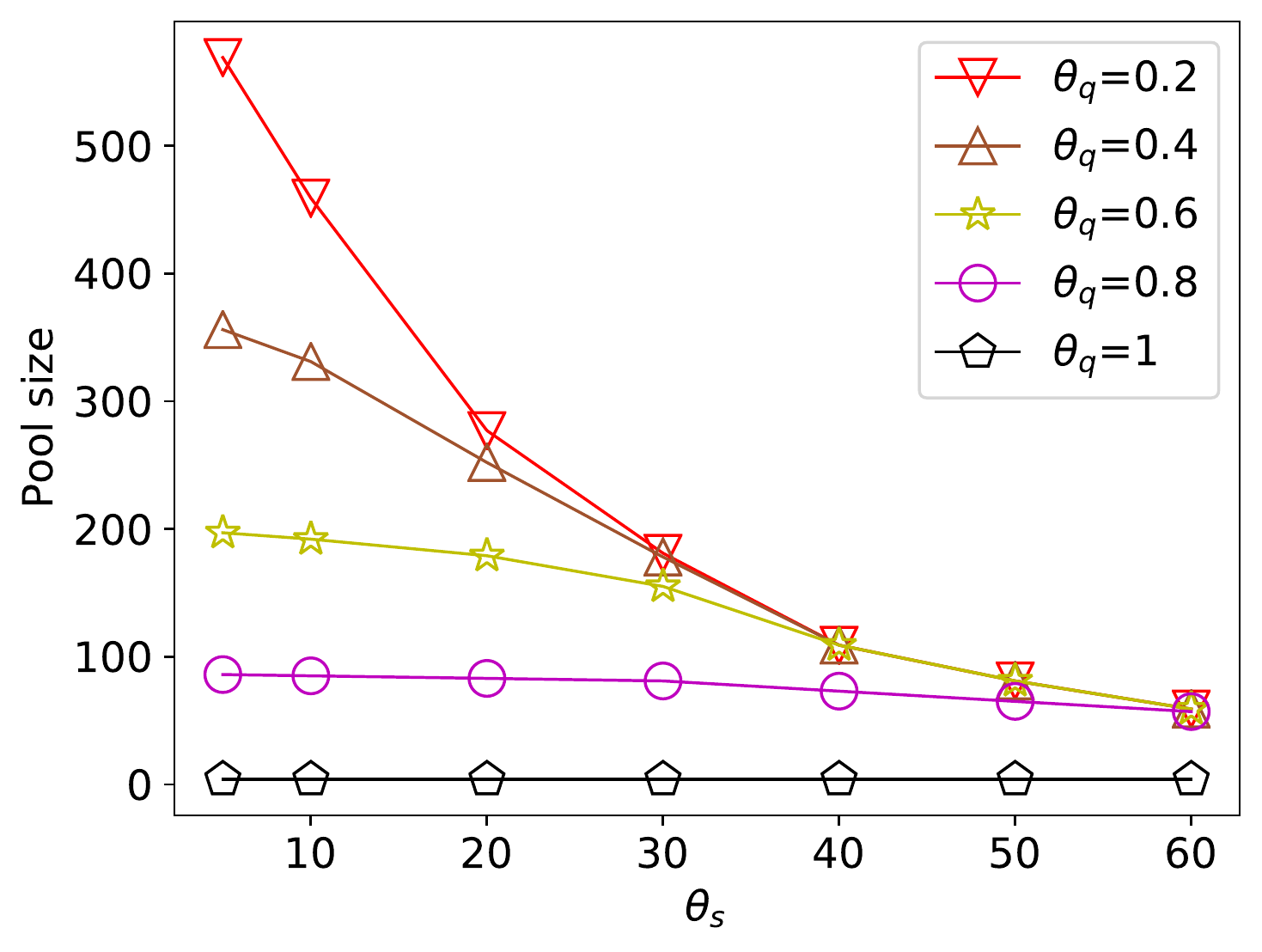}\label{fig:poolsize_fb}}	
	\subfigure[Git]{\includegraphics[width=0.48\columnwidth]{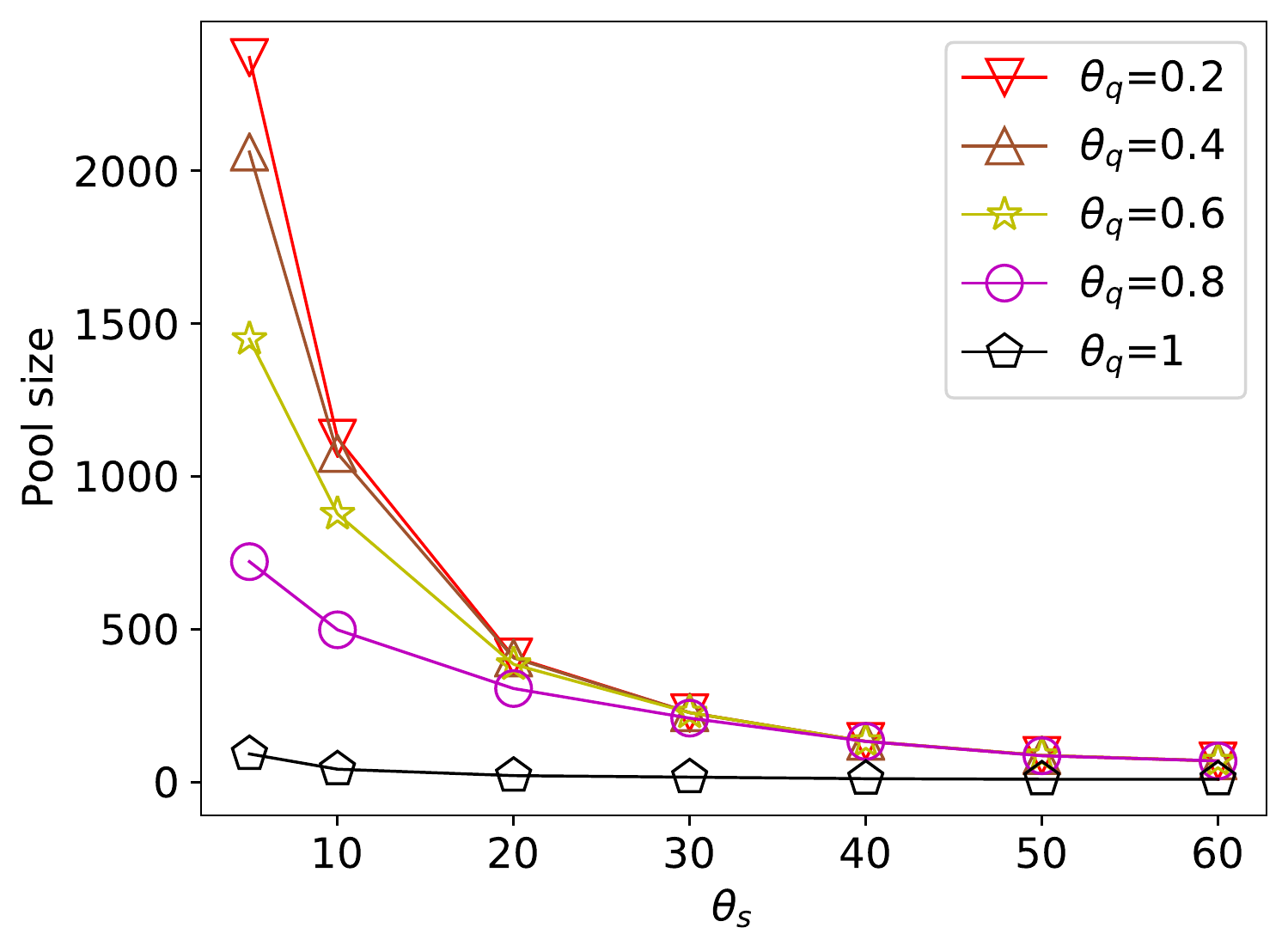}\label{fig:poolsize_git}}\\
	\caption{Pool size with different parameter settings}
	\label{fig:poolsize}
\end{figure}

Figures \ref{fig:poolsize_fb} and \ref{fig:poolsize_git} demonstrate the impact on influencer pool size by varying both parameters in the Ego-Facebook and Git datasets, respectively. It is evident that in both figures, the influencer pool size shows a downside trend with the increase of $\theta_s$ or $\theta_q$. However, in the Git dataset, the influencer pool size is more sensitive to $\theta_s$ than that of the Ego-Facebook dataset. This is because the node connectivity in Git appears to turn out to be much sparser than Ego-Facebook. 

Third, we investigate how the influencer pool size impacts the ABEM's performance. It is important to strike a balance between efficiency and effectiveness. Specifically, a larger influencer pool size enables ABEM to find out a better solution, but with less efficiency due to the large scope. ABEM can converge more rapidly with a smaller influencer pool but may not yield a better solution. This is due to the reason that potential influencers may be filtered out when decreasing the size of influencer pool.

\begin{figure}[!htb]
	\centering
	\subfigure[Ego-Facebook]{\includegraphics[width=0.48\columnwidth]{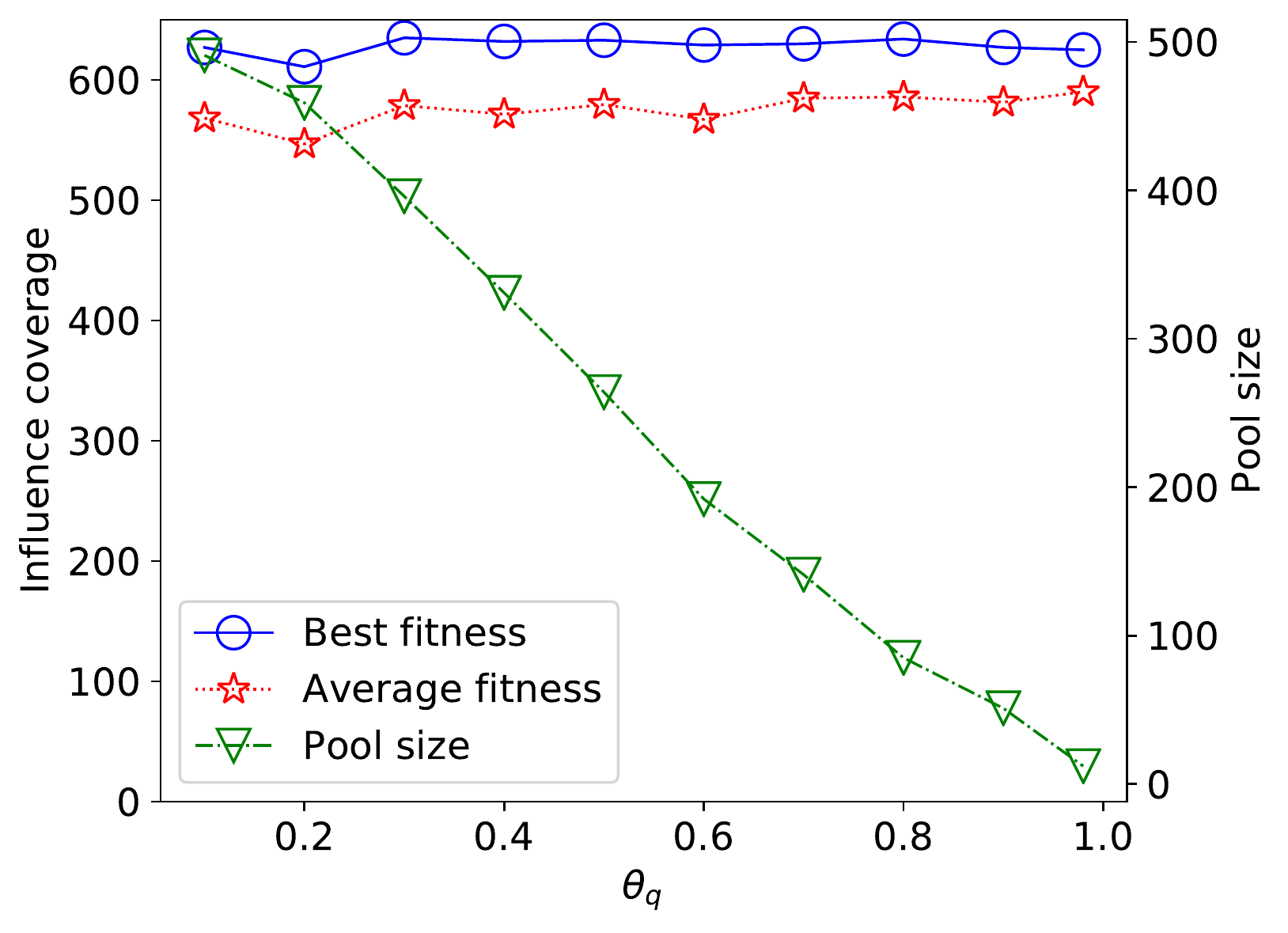}\label{fig:poolsize_fb_influence}}	
	\subfigure[Git]{\includegraphics[width=0.48\columnwidth]{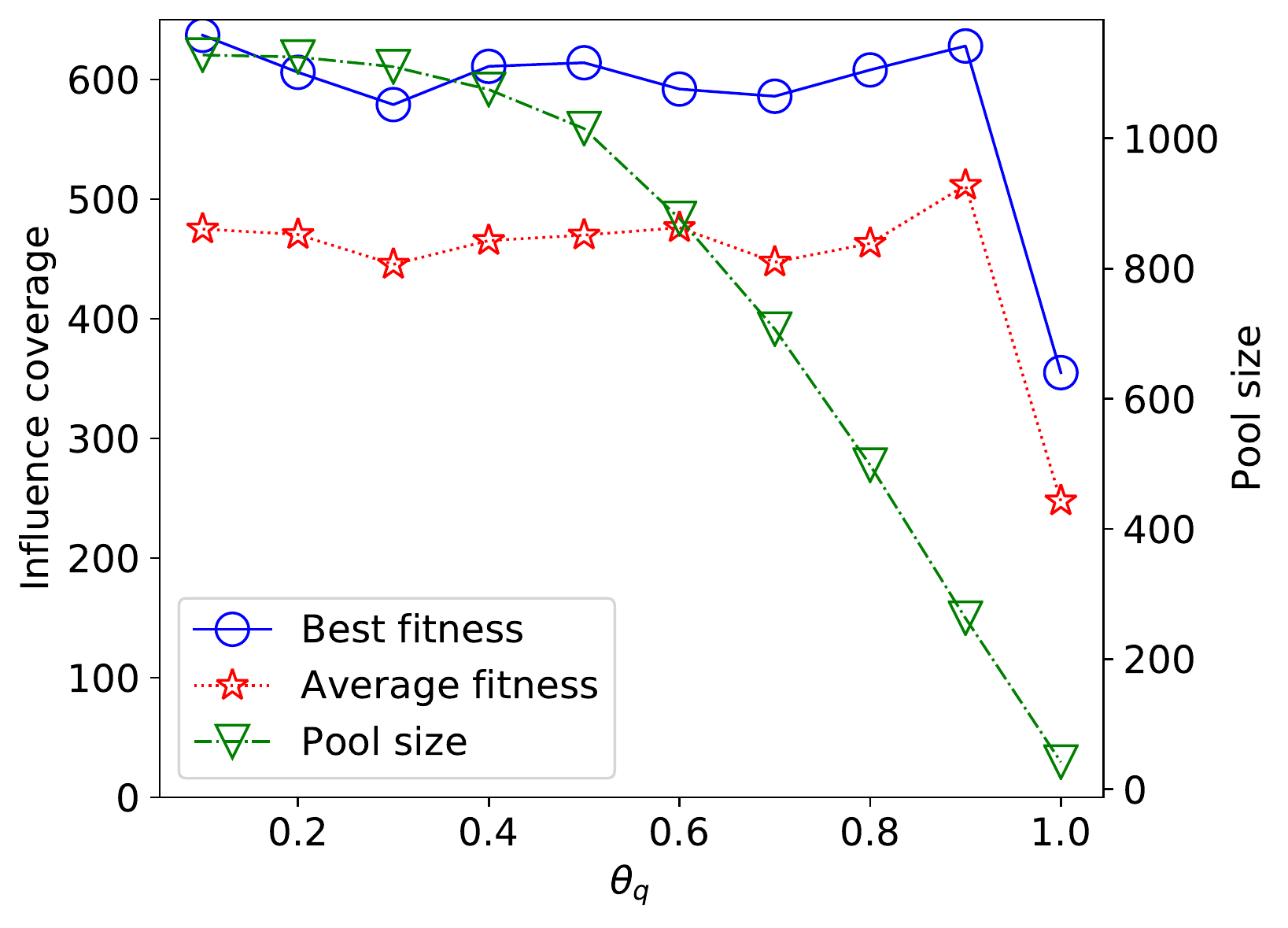}\label{fig:poolsize_git_influence}}\\
	\caption{Pool size analysis}
	\label{fig:poolsize_influence}
\end{figure}

Figures \ref{fig:poolsize_fb_influence} and \ref{fig:poolsize_git_influence} demonstrate the influencer pool size analysis by using two datasets. In the Ego-Facebook dataset, no significant performance improvement can be observed when decreasing the influencer pool size. In other words, the smallest size enables the ABEM to yield almost the same performance as that of a larger pool size. This phenomenon also implies ABEM will carry out a similar seed set as that of the degree-based selection in the Ego-Facebook network. This is also consistent with the results of Experiment 2. By contrast, Figure \ref{fig:poolsize_git_influence} reveals a different pattern. The best fitness shows a steady trend until $\theta_q $ reaches 0.9. Starting from this point, both best and average fitness drop dramatically. This is because potential influencers are filtered out due to narrowing down the searching scope. Therefore, by considering both efficiency and effectiveness, ABEM will adopt $\theta_q =0.9 $ with an influencer pool size of 600. In this case, ABEM will definitely outperform degree-based selection, which is also consistent with the results of Experiment 2. 

\subsection{Experiment 4: Influence maximization in dynamic networks}
\label{sec:exp4}
Experiment 4 aims to evaluate the performance of ABEM in dynamic online social networks. This experiment also explicitly demonstrates the proven adaptability of ABEM, namely, continuously updating the identified seed sets in a changing environment and adapting the solutions based on past experiences.

\begin{table*}
	\caption{Flixster quarterly network}
	\label{tab:flixster_quaterly_1}
	\begin{adjustbox}{width=1\textwidth}
		\small
		\begin{tabular}{cllllllllllll}
			\toprule
			Snapshot ID&1&2&3&4&5&6&7&8&9&10&11&12\\
			\midrule
			Quarter&2006 Q1&2006 Q2&2006 Q3&2006 Q4&2007 Q1&2007 Q2&2007 Q3&2007 Q4&2008 Q1&2008 Q2&2008 Q3&2008 Q4\\
			No. of nodes & 463&564&825&1287&2958&2993&2051&1280&1387&1529&1615&1353\\
			No. of links & 1050&1614&2146&4027&8965&8611&5789&3517&4252&4758&4610&3907\\
			Average degree & 
			4.54&5.72&5.20&6.26&6.06&5.75&5.65&5.50&6.13&6.22&5.71&5.78\\
			\bottomrule
		\end{tabular}
	\end{adjustbox}
\end{table*}

In this experiment, the dynamic environment is simulated by using 12 consecutive quarters' transactions from the Flixster dataset, ranging from 2006 to 2008 \cite{jamali2010amatrix}. The statistics of the dataset is listed in Table \ref{tab:flixster_quaterly_1}. Since the size of some snapshots appears small, we give $k=5$, $\theta_s = 2$, and $\theta_q = 0.7$. Five seeding algorithms, i.e., Greedy, Degree, DDH, GA, and GA with influence pool, are utilized as the counterparts. In the influence maximization problem, the greedy algorithm is recognized as one of the strongest baselines \cite{chen2009efficient,kempe2003maximizing}, which is also reflected in Experiment 2. On top of that, we list the assumptions as follows. 

\begin{figure*}[!htb]
	\centering
	\subfigure[k=5]{\includegraphics[width=0.32\textwidth]{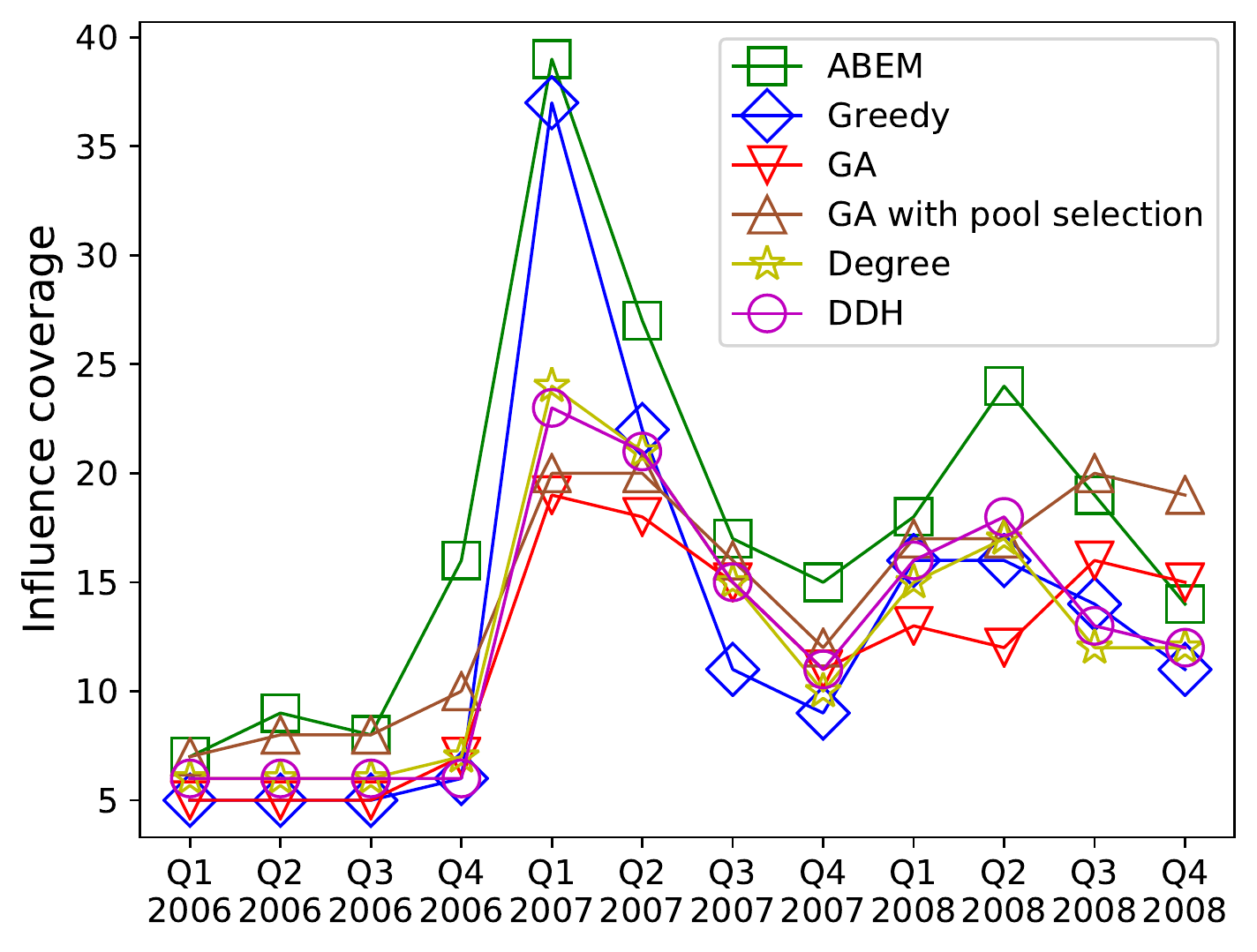}}	
	\subfigure[k=10]{\includegraphics[width=0.32\textwidth]{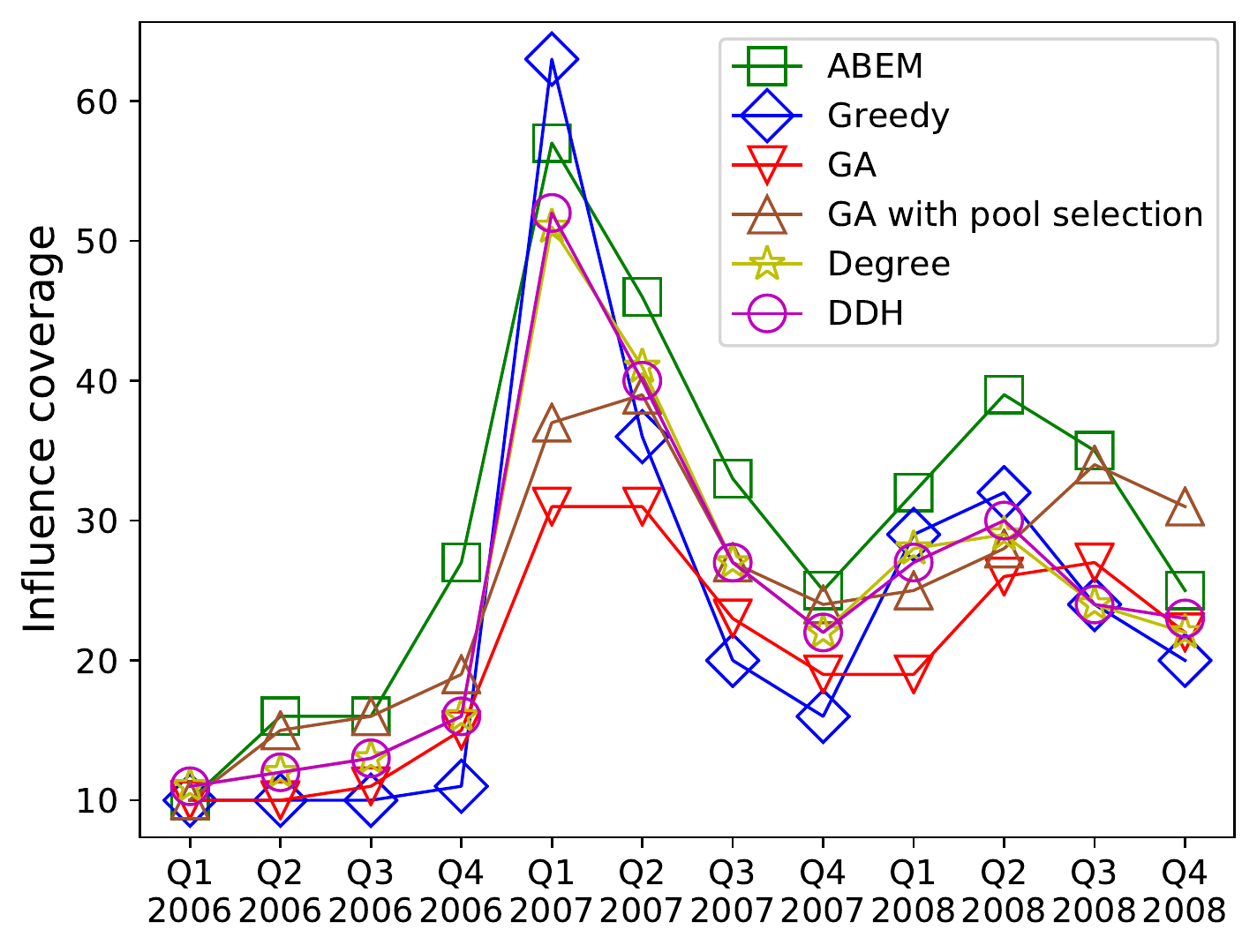}}	
	\subfigure[k=15]{\includegraphics[width=0.32\textwidth]{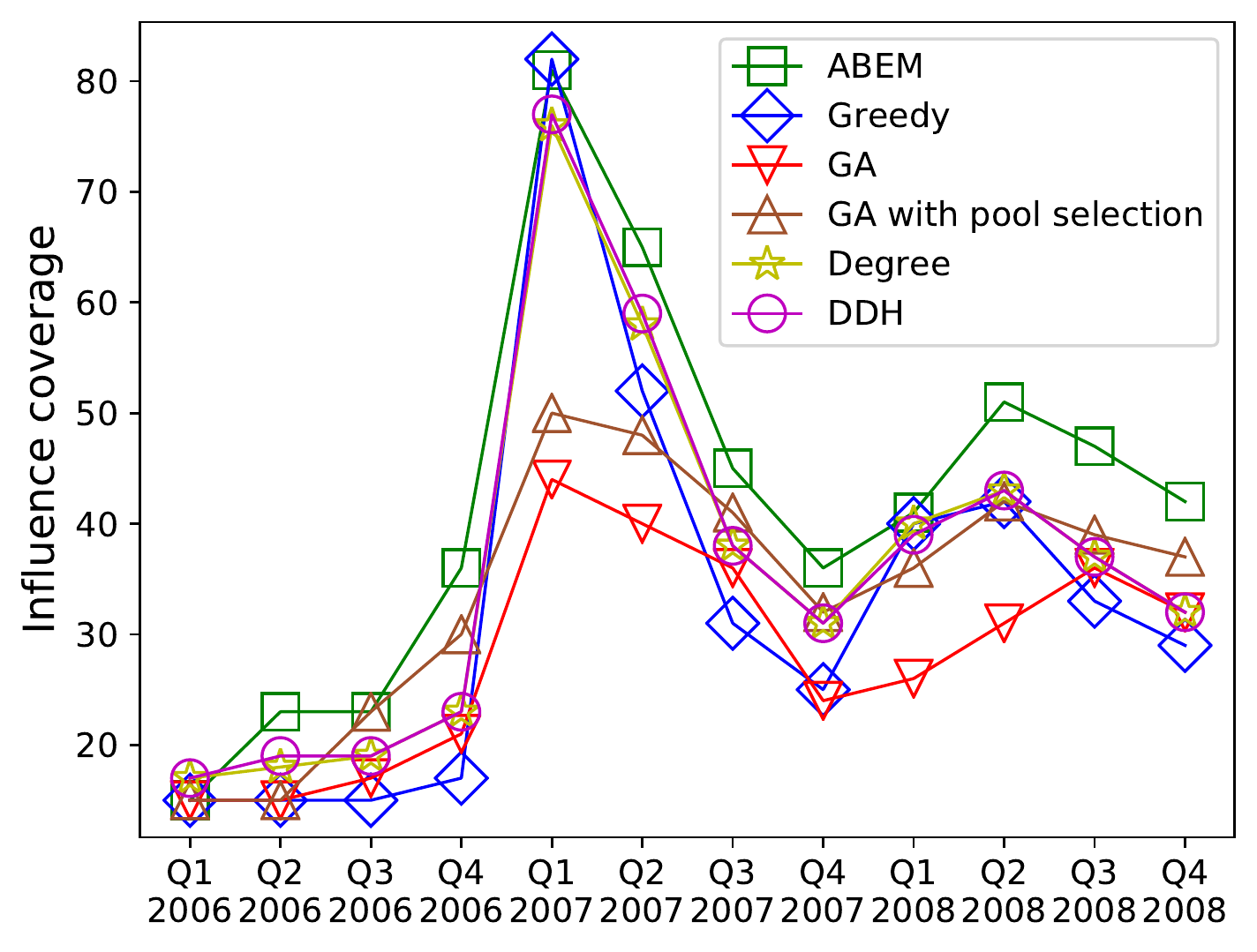}}\\
	\caption{ABEM performance in Flixster quarterly network}
	\label{fig:abem_quarterly}
\end{figure*}

\begin{itemize}
	\item A user joins the network when giving the first rating, and quits after the last rating. A user only can be influenced when he or she appears an active user in the social network. 
	\item When a user joins the network, the corresponding relationships are established immediately. Likewise, the associated links are removed if the user quits the network. 
	\item The greedy algorithm recalibrates the selected seed set on an annual basis. This is because the greedy algorithm is not scalable for large scale networks. It is unrealistic to launch the greedy algorithm frequently. 
	\item The Degree and DDH reselect influencers every four quarters as well. This is because these heuristic algorithms require the entire network topology. It is unrealistic to rank all of the users' degree in large-scale networks frequently.
\end{itemize}
First, we compare the influence coverage of ABEM against the other baselines with different seed set sizes, i.e., $k=5$, $k=10$, and $k=15$. Figure \ref{fig:abem_quarterly} illustrates the experimental results using 12 consecutive quarters, where the network topology, including both nodes and links, evolves over time. As can be seen from the figure that ABEM outperforms the classic seeding algorithms, implying that the greedy algorithm losses the advantages in a changing environment if without any calibration. By contrast, ABEM adapts the solutions over time, and this feature stems from its internal design, i.e., a hybrid of multi-agent systems and evolutionary computation. Furthermore, ABEM outperforms the other two evolutionary approaches, i.e., GA and GA with influence pool algorithms, in dynamic networks. This is because ABEM leverages the influencer pool for initialization and re-calibration, which gives ABEM a better starting point and a higher chance to fast converge to an optimal solution.

\begin{figure*}[!htb]
	\centering
	\subfigure[k=5]{\includegraphics[width=0.32\textwidth]{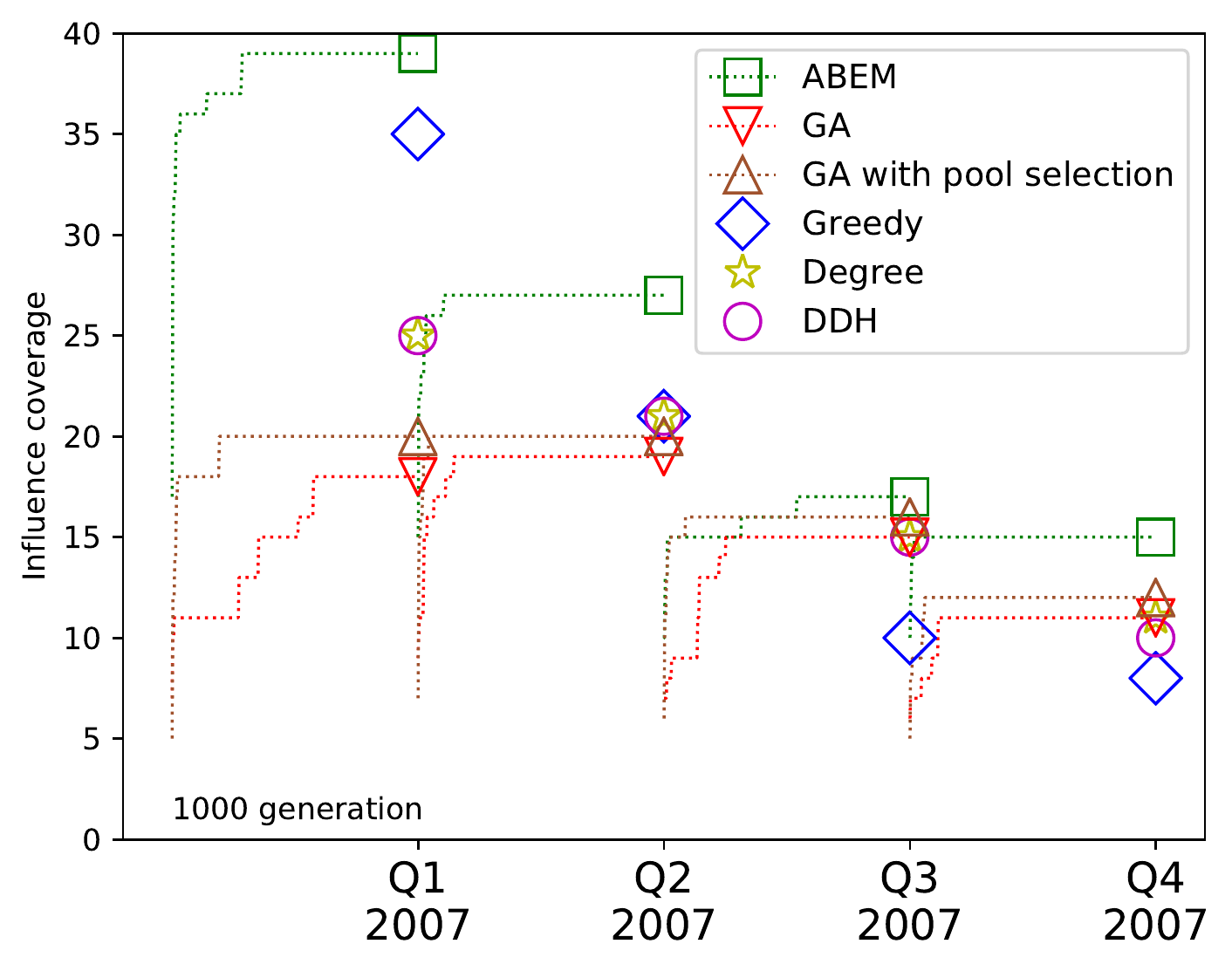}}
	\subfigure[k=10]{\includegraphics[width=0.32\textwidth]{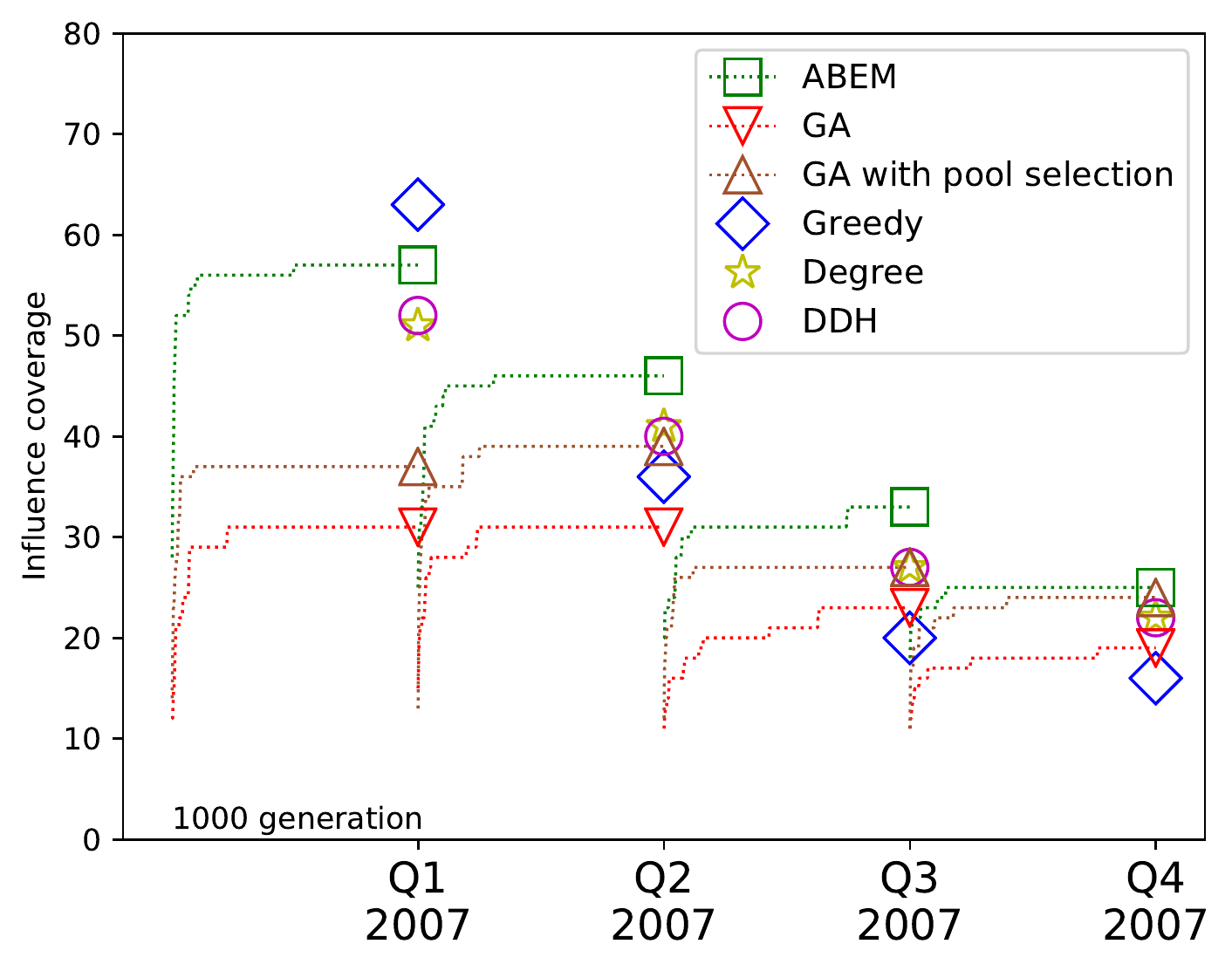}}
	\subfigure[k=15]{\includegraphics[width=0.32\textwidth]{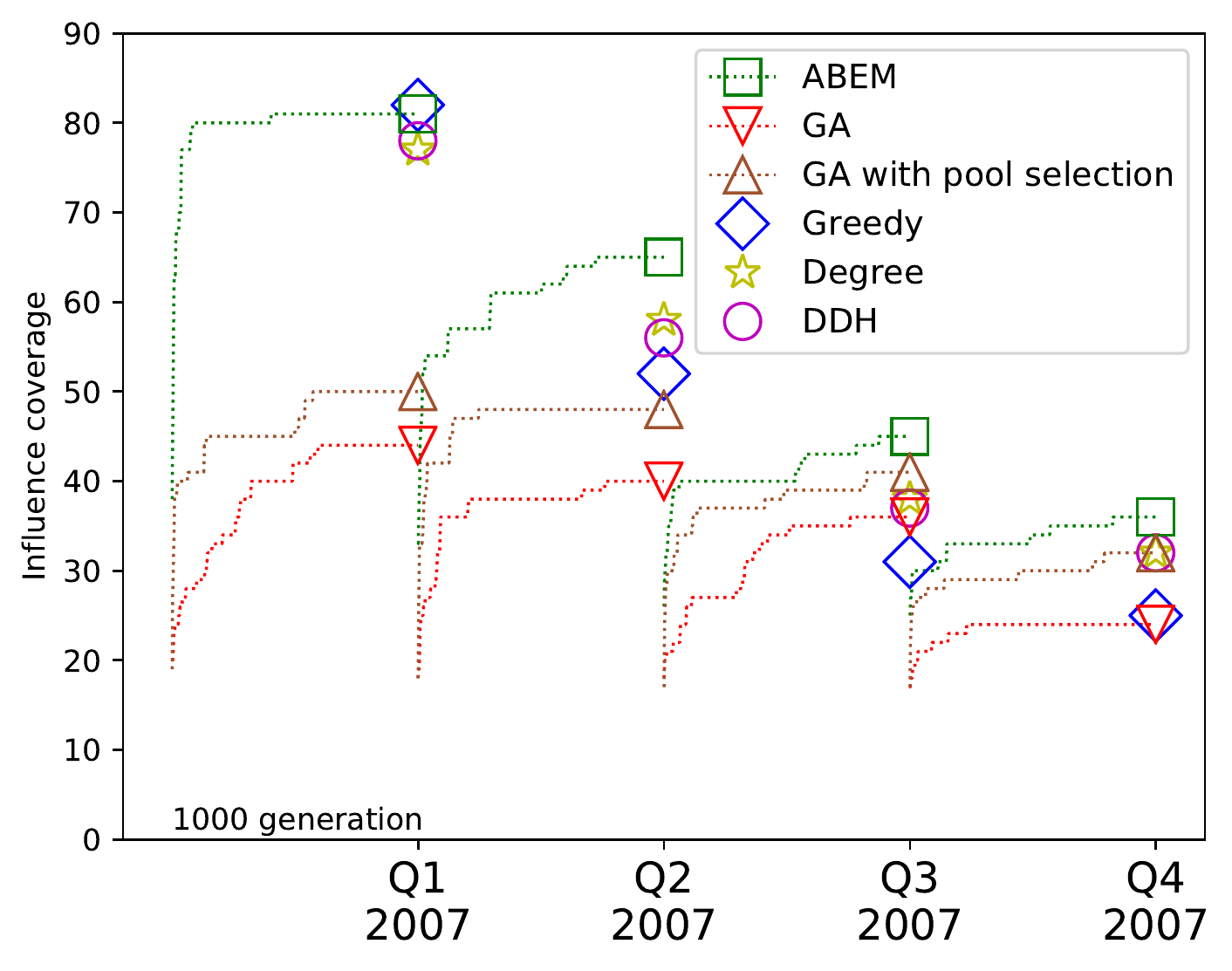}}\\
	\caption{ABEM performance in dynamic environment}
	\label{fig:abem_timestep}
\end{figure*}


Second, we further investigate how ABEM adapts the seed set in dynamic networks by using the same dataset, where four consecutive quarters from Flixster, i.e., from 2007 Q1 to 2007 Q4, are selected. The evolutionary process of ABEM is demonstrated in Figure \ref{fig:abem_timestep}, where the x-axis represents discrete time steps, i.e., network snapshots, and the y-axis denotes the influence coverage produced by the algorithm. 1000 units are allocated between any two consecutive quarters, and each unit presents a generation produced by  evolutionary algorithms. We have selected a seed set size of $k=5$, $k=10$, and $k=15$ for the exploration. As can be seen from these figures that in Q2 2007, ABEM quickly reaches an optimal solution merely within 500 generations. When the network evolves, in Q2 2007, the performance of ABEM drops but climbing up quickly only after a few generations. This is because the existing potential influencers are retained in the influencer pool and the solution can be adapted rapidly. In Q3 2007, ABEM requires a greater number of generations to coverage. The reason is dramatic variations occur in the network at this point, many existing influencers need to be replaced within the population. Therefore, it can be proved that ABEM has great adaptability to handle the dynamics of social network efficiently. 

Third, we further validate the adaptive capabilities of ABEM with different parameter settings. Recall that the influencer pool of ABEM works as a common knowledge repository and scopes the problem searching space. Therefore, influence pool significantly affects the performance of ABEM. In this experiment, we observe the outcome by varying the degree threshold $\theta_s$ and influence quartile threshold $\theta_q$ of the influencer pool. 

\begin{figure}[!htb]
	\center
	\includegraphics[width=0.5\textwidth]{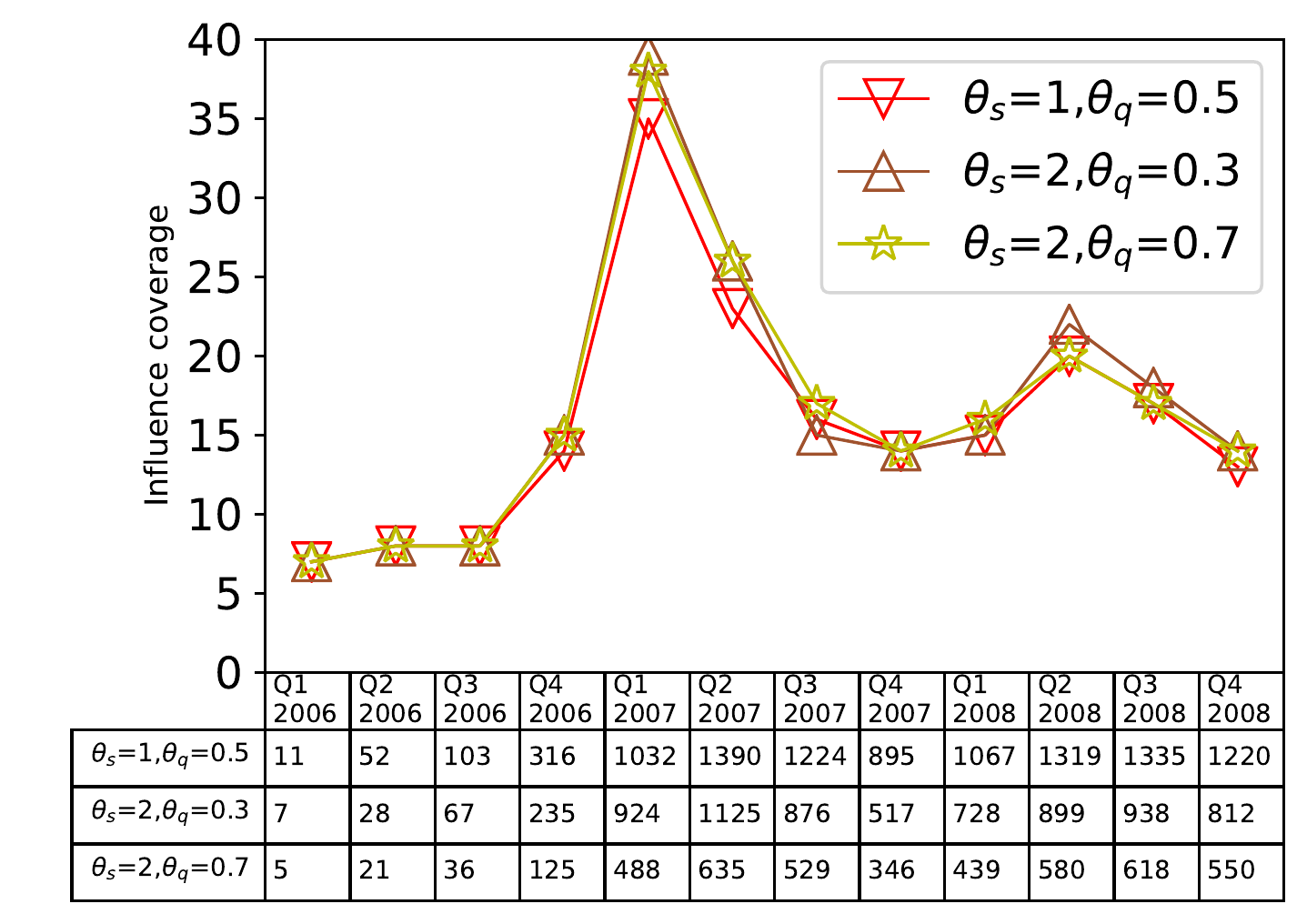}
	\caption{ABEM performance with different pool size}
	\label{fig:exp4_abem_poolsize}
\end{figure}

Figure \ref{fig:exp4_abem_poolsize} shows the influence coverage with three different settings over 12 quarters: (1) $\theta_s=1$ and $\theta_q=0.5$ (2) $\theta_s=2$ and $\theta_q=0.3$ (3) $\theta_s=2$ and $\theta_q=0.7$, with a seed set size of 5. The table of Figure \ref{fig:exp4_abem_poolsize} depicts the detailed outcome based on various settings. 

Not much difference can be observed when the network size is small, i.e., from Q1 2006 to Q3 2006. However, with the increase of network scale, the pool size shows a great impact on influence coverage. A larger pool size ($\theta_s=1$ and $\theta_q=0.5$) leads a relatively lower performance. This is because the searching space of ABEM is expanded and it requires more generations to reach an optimal solution. By contrast, given a small pool size (i.e., $\theta_s=2$ and $\theta_q=0.7$), ABEM yields better performance than those with different settings in most of the snapshots but demonstrates a relatively weak performance in 2008 Q2 and Q3. This suggests that the over shrink of the influencer pool inevitably filters out some potential influencers.

\section{Conclusion} \label{sec:conclusion}
In this paper, a novel agent-based evolutionary approach, i.e., ABEM, is proposed to mine influencers in online social networks. We elaborate the proposed approach, including algorithms, mining process and each component of ABEM in details. We also clarify the major capabilities of ABEM, i.e., handling large-scale networks and dynamic environments. The former relies on the agent-based modelling, where the major computational cost can be distributed to the individual agent in ABEM. The behavioural outcome provides a reasonable searching scope for ABEM. The latter is handled by the proposed algorithms, which can retain the existing potential influencers and modify part of the solutions. Extensive experiments are conducted to evaluate the performance and capability of ABEM, including convergence analysis, influence maximization in both static and dynamic networks, and parameter analysis. The experimental results demonstrate that ABEM not only outperforms the state-of-the-art algorithms in mining influencers in static networks, but also can be applied to a large-scale and dynamic environment.

In the future, we plan to further improve ABEM by employing some heuristics, which can expedite the convergence speed in a changing environment. Furthermore, we will develop an enhanced version of ABEM to fit a more sophisticated influence diffusion process with the consideration of context.

\newpage

\bibliographystyle{ACM-Reference-Format}
\bibliography{manuscript}

\end{document}